\documentclass[%
reprint,
 amsmath,amssymb,
 aps,
]{revtex4-2}

\usepackage{graphicx}
\usepackage{hyperref}
\usepackage{dcolumn}
\usepackage{bm}
\usepackage{color}
\usepackage{braket}

\DeclareUnicodeCharacter{2212}{-}

\begin{document}

\preprint{APS/123-QED}

\title{Double-dome Unconventional Superconductivity in Twisted Trilayer Graphene}

\author{Zekang Zhou$^{1}$, Jin Jiang$^{1}$, Paritosh Karnatak$^{2}$, Ziwei Wang$^{3}$, Glenn Wagner$^{4}$, Kenji Watanabe$^{5,6}$, Takashi Taniguchi$^{5,6}$, Christian Sch\"{o}nenberger$^{2,7}$, S. A. Parameswaran$^{3,8}$, Steven H. Simon$^{3}$, Mitali Banerjee$^{1,}$}

\email{mitali.banerjee@epfl.ch}

\affiliation{$^1$Institute of Physics, Ecole Polytechnique Fédérale de Lausanne, CH-1015 Lausanne, Switzerland\\$^2$Department of Physics, University of Basel, Klingelbergstrasse 82, CH-4056 Basel, Switzerland\\$^3$Rudolf Peierls Centre for Theoretical Physics, Parks Road, Oxford, OX1 3PU, UK\\$^4$Department of Physics, University of Zurich, Winterthurerstrasse 190, 8057 Zurich, Switzerland\\$^5$Research Center for Functional Materials, National Institute for Material Science, 1-1 Namiki, Tsukuba 305-0044, Japan\\$^6$International Center for Materials Nanoarchitectonics,
National Institute for Material Science, 1-1 Namiki, Tsukuba 305-0044, Japan\\$^7$Swiss Nanoscience Institute, University of Basel, Klingelbergstrasse 82, CH-4056 Basel, Switzerland\\$^8$Max Planck Institute for the Physics of Complex Systems, Nöthnitzer Str. 38, 01187 Dresden, Germany}

\begin{abstract}
Graphene moiré systems are ideal environments for investigating complex phase diagrams and gaining fundamental insights into the mechanisms underlying exotic states of matter, as they permit controlled manipulation of electronic properties. Magic-angle twisted trilayer graphene (MATTG) has emerged as a key platform to explore moiré superconductivity, owing to the robustness of its superconducting order and the displacement-field tunability of its energy bands. Recent measurements strongly suggest that superconductivity in MATTG is unconventional. Here, we report the first direct observation of double-dome superconductivity in MATTG. The temperature, magnetic field, and bias current dependence of the superconductivity of doped holes collectively show that it is significantly suppressed near moiré filling $\nu^* = -2.6$, leading to a double dome in the phase diagram within a finite window of the displacement field. The temperature dependence of the normal-state resistance and the $I-V$ curves straddling $\nu^*$  are suggestive of a phase transition and the potentially distinct nature of superconductivity in the two domes. Hartree-Fock calculations incorporating mild strain yield an incommensurate Kekulé spiral state whose effective spin polarization peaks in the regime where superconductivity is suppressed in experiments. This allows us to draw conclusions about the normal state as well as the unconventional nature of the superconducting order parameter. 
\end{abstract}

\maketitle

\section*{Introduction}
Double-dome superconductivity has been found in various systems such as cuprates~\cite{koike1991effects,grissonnanche2014direct}, heavy fermion compounds
~\cite{yuan2003observation,grosche2000anomalous}, iron chalcogenides~\cite{zhang2017observation} and kagome lattice materials~\cite{zhang2021pressure}. The appearance of a double dome in a superconducting phase diagram usually signals unconventional superconductivity and is associated with intertwined orders, quantum criticality, and non-Fermi liquid physics~\cite{das2016two}. In most systems mentioned above, the origin of double-dome superconductivity 
cannot be explained by conventional BCS theory and is typically attributed to electron correlations. Recently, twisted graphene moiré superlattices, whose nearly flat bands enhance the role of Coulomb interactions, have emerged as exciting platforms to explore quantum phases such as correlated insulators~\cite{cao2018correlated}, superconductors~\cite{cao2018unconventional,yankowitz2019tuning,park2021tunable,hao2021electric,zhang2022promotion,park2022robust}, ferromagnets~\cite{sharpe2019emergent,serlin2020intrinsic,lin2022spin} and (fractional) Chern insulators~\cite{nuckolls2020strongly,das2021symmetry,saito2021hofstadter,wu2021chern,park2021flavour,choi2021correlation,pierce2021unconventional,xie2021fractional}. Among these, superconductivity has attracted much attention and has already been extensively studied in magic angle twisted bilayer graphene (MATBG)~\cite{cao2018unconventional} and twisted multilayer graphene~\cite{park2021tunable,hao2021electric,park2022robust,zhang2022promotion}. Importantly, aspects of the superconducting order in these systems --- such as its nematicity~\cite{cao2021nematicity}, re-entrant behaviour~\cite{cao2021pauli}, and signatures of nodal pairing~\cite{kim2022evidence,oh2021evidence} --- all suggest that it may be unconventional, highlighting the role of electron interactions. Here, we report the direct observation of  double-dome superconductivity in MATTG tunable by the application of an electrical displacement field, and differentiate the characteristics of the two domes through electronic transport measurements. Our findings, bolstered by numerical calculations, confirm unconventional superconductivity in MATTG and provide valuable insights toward understanding the mechanism of superconductivity in this material. 

\section*{Superconductivity and correlated states}

\begin{figure*}
\centering
\includegraphics[width= 0.9327\textwidth]{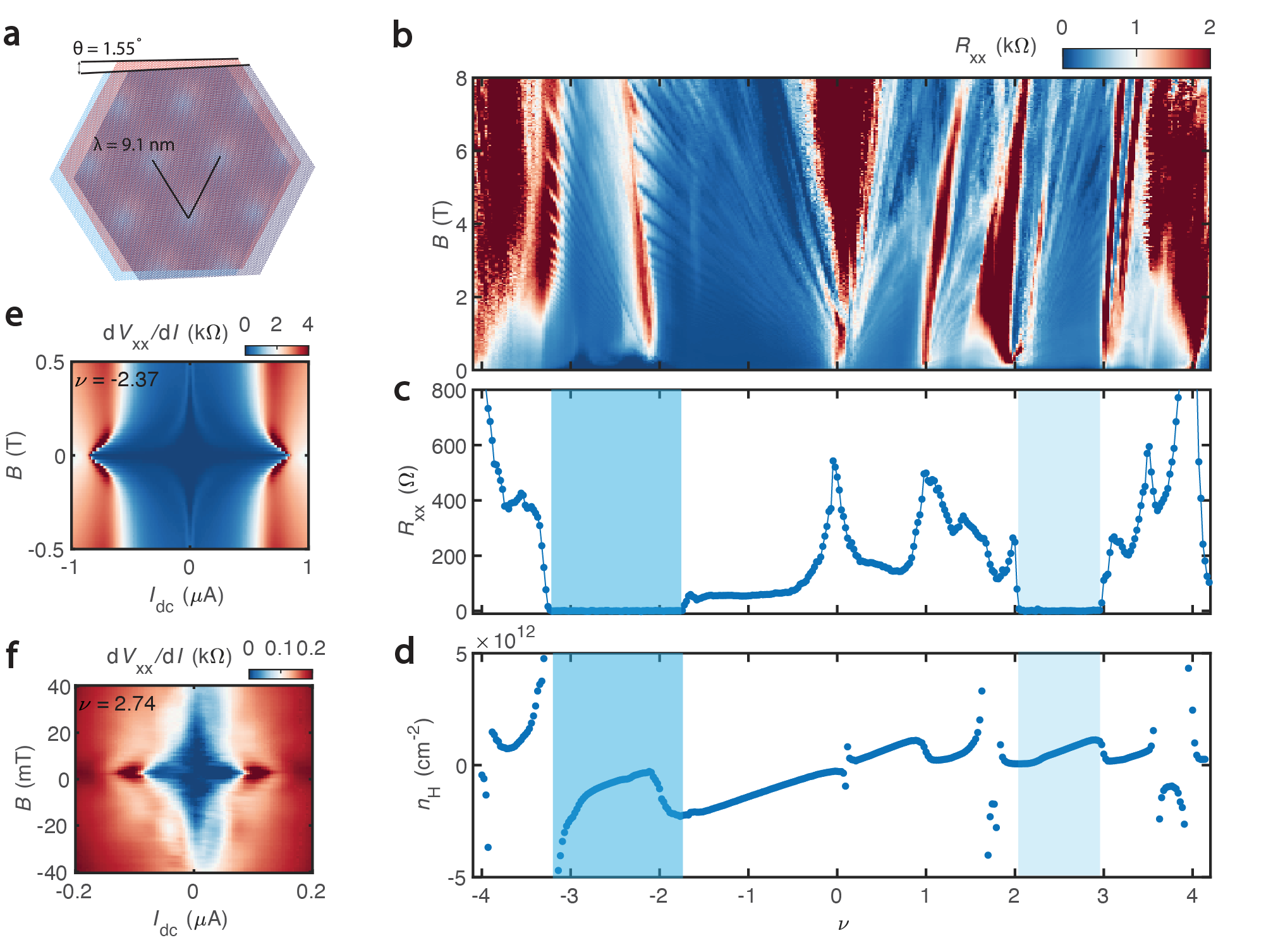}
\caption{\textbf{Superconducting phases in Twisted Trilayer Graphene (\textbf{$T$} = 100 mK).}  
\textbf{a}, Schematic showing the moiré pattern with $9.1$~nm moiré length in MATTG in which the alternating twist angle $\theta_{12}=1.55^{\circ}$ and $\theta_{23}=-1.55^{\circ}$. \textbf{b}, Landau fan diagram shown by longitudinal resistance $R_{xx}$. Correlated states can be seen at $\nu=1, \pm2, 3$. Landau levels also start from these correlated states in addition to the charge neutral point (CNP) and moiré band full filling. The non-zero Chern number $C=\pm2$ of these correlated states comes from Landau levels of the Dirac band. \textbf{c}, linecut of $R_{xx}$ in zero magnetic field. The superconducting phase appears near $\nu=−2$ and $\nu=−3$ in the hole side and appears on the electron side between $\nu=2$ and $\nu=3$ (the light blue region). \textbf{d}, Hall density $n_H$ extracted from $R_{xy}=B/ne$. $n_H$ shows `gap/Dirac' behavior at $\nu = 0, \pm4$, `reset' behavior at $\nu = 1, \pm2, 3$ and `vHS' between $\nu = -4, -3$, between $\nu = 1, 2$, between $\nu = 3, 4$~~\cite{park2021tunable}. The hole side superconducting phase is bounded by the `vHS' and `reset' point while the electron side superconducting phase is bounded by two `reset' points, which is in agreement with previous works~\cite{park2021tunable,hao2021electric}. \textbf{e, f}, $dV_{xx}/dI$ versus $I_{dc}$ and $B$ in the hole side superconducting phase (\textbf{e}, $\nu = -2.37$) and in the electron side superconducting phase (\textbf{f}, $\nu = 2.74$). The hole side superconducting phase is much stronger than the electron side. No Fraunhofer interference pattern is observed either on the hole side or the electron side.}
\label{fig1}
\end{figure*}

\begin{figure*}
\centering
\includegraphics[width= 0.84\textwidth]{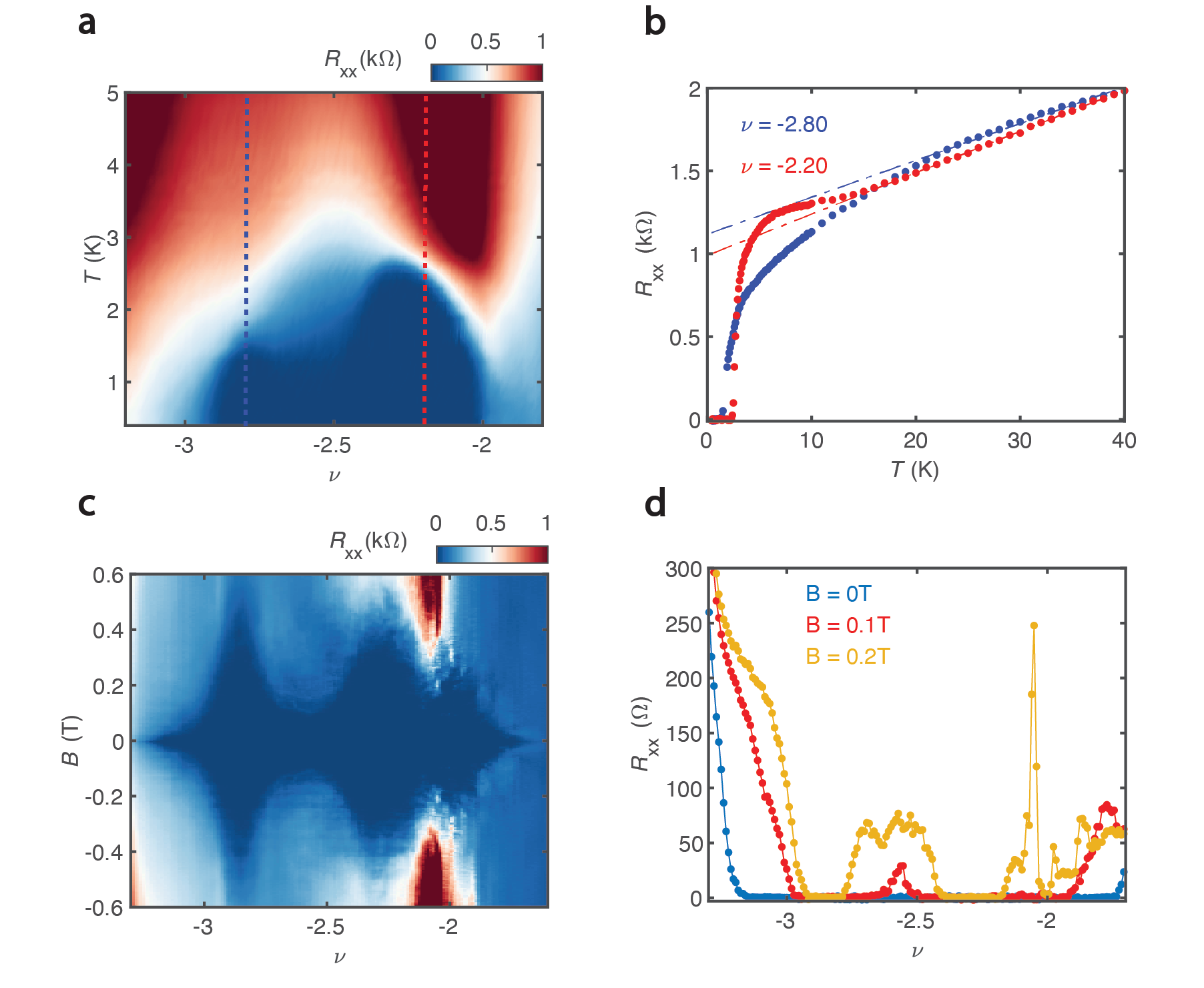}
\caption{\textbf{Temperature and magnetic field dependence show double dome superconductivity in MATTG. ($T$ = 100 mK except temperature dependence)} \textbf{a}, Temperature dependence of $R_{xx}$ in the hole side superconducting regime.  The superconducting phase is strongly suppressed near $\nu^*=-2.6$. \textbf{b}, line cuts of $R_{xx}$ in \textbf{a} when $\nu=-2.2$ and $\nu=-2.8$. Dashed lines are the linear fit of $R_{xx}$ versus $T$ in the high-temperature region. \textbf{c}, The magnetic field dependence of $R_{xx}$ also shows a clear feature of double dome superconductivity. The left and right dome's superconductivity are very robust, with a maximum critical magnetic field larger than 300 mT. Near $\nu^*=-2.6$, superconductivity is sufficiently weakened. \textbf{d}, Line-cuts from \textbf{c} at three different magnetic fields. At zero magnetic field, the superconducting region extends below $\nu=-3$ and above $\nu=-2$. When $B$ = 0.1T, the superconducting regime shrinks from both sides, and a small resistance peak appears near $\nu^*=-2.6$. When $B$ = 0.2T, the resistance peak near $\nu^*=-2.6$ becomes enhanced, and the $\nu=-2$ correlated state appears. 
}
\label{fig2}
\end{figure*}

\begin{figure*}
\centering
\includegraphics[width= 1\textwidth]{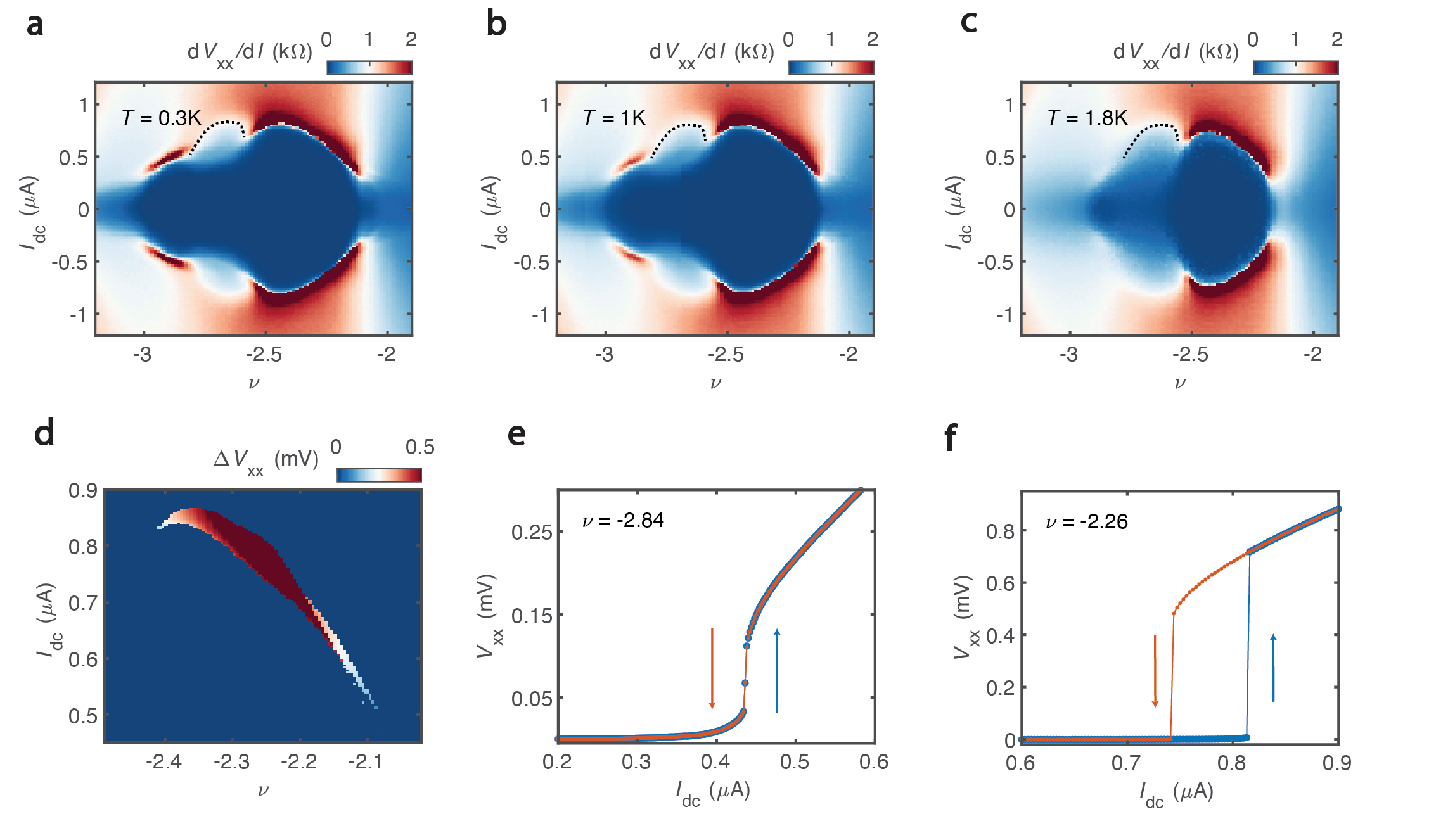}
\caption{\textbf{Bias current dependence shows two dome superconductivity and $I-V$ curve hysteresis. (100mK)}  \textbf{a, b, c}, $dV_{xx}/dI$ as a function of $\nu$ and $I_{dc}$ at 0.3 K (\textbf{a}), 1 K (\textbf{b}) and 1.8 K (\textbf{c}). The phenomenon of double dome superconductivity remains evident. Clear critical current features, denoted as $I_c$, are observable within the ranges of $-3<\nu<-2.8$ and $-2.5<\nu<-2.1$. 
At temperature 1.8 K, the superconducting feature of the left dome nearly disappears, whereas the right dome retains its superconducting characteristics, appearing as an isolated dome. \textbf{d}, The hysteresis of the $I-V$ curve in the right dome, denoted as $\Delta V_{xx} = (V_{xx}(\leftarrow)-V_{xx}(\rightarrow))$, varies as a function of filling and d.c. bias current. Notably, a distinct hysteresis behavior is observed in the $-2.4<\nu<-2.1$ range near the critical current. \textbf{e, f}, $V_{xx}(\leftarrow)$(red) and $V_{xx} (\rightarrow)$(blue) at filling $\nu=-2.84$ (\textbf{e}) and $\nu=-2.6$ (\textbf{f}). The left dome exhibits no hysteresis, whereas the right dome demonstrates a pronounced hysteresis.}
\label{fig3}
\end{figure*}

Magic angle twisted trilayer graphene (MATTG) comprises three layers of graphene, with the first and third layers having the same orientation and the central layer twisted by approximately $1.55^\circ$~\cite{khalaf2019magic}. This configuration is mirror-symmetric for zero displacement field and exhibits a moiré pattern with a wavelength of around $9.1$ nm, as illustrated in FIG. \ref{fig1}a. Similar to MATBG, the low energy physics in MATTG is dominated by flat bands. However, a distinguishing feature of MATTG, as opposed to MATBG, is the presence of a Dirac band alongside the flat bands, that is decoupled from them at zero displacement field. Applying a displacement field  to MATTG breaks mirror symmetry and hybridizes flat and Dirac bands, thereby modifying the band structure~\cite{khalaf2019magic} and providing an elegant route to explore the interplay between various competing or intertwined states. We have fabricated multiple TTG devices close to the magic angle for this work (details of the stacking and fabrication process are contained in Methods). Similar to MATBG,  MATTG also exhibits a rich phase diagram hosting several quantum phases whose explanation lies beyond single-particle physics. FIG. \ref{fig1}b shows longitudinal resistance $R_{xx}$ as a function of the magnetic field $B$ and bottom gate voltage $V_{bg}$, which is converted into moiré filling factor $\nu=n/n_0$ where $n$ and $n_0$ are the total electron density and that needed to fill a single moiré band, respectively. Apart from the resistance peaks observed at $\nu = \pm4$, which stem from the band gap of the flat band, and at $\nu = 0$, corresponding to semi-metallic behavior at neutrality, additional resistance peaks are evident at $\nu = 1, \pm2, 3$. These cannot be accounted for solely by the non-interacting band structure and are correlated states that form due to the strong interaction-induced symmetry breaking within the flat bands. For $B\neq 0$, Landau levels also originate from these correlated states, indicating that the Fermi surface reconstructs at these moiré fillings. Furthermore, the correlated states exhibit a slope in the $\nu-B$ plane that translates via the Streda formula $d\nu/dB = C e/h n_0$ to a Chern number $\lvert C \lvert=2$. However, unlike in MATBG where the Chern number arises solely from the topology of correlated states in the flat bands~\cite{wu2021chern,nuckolls2020strongly,choi2021correlation,saito2021hofstadter,das2021symmetry}, in MATTG it can be ascribed to the Dirac band Landau levels~\cite{shen2023dirac}, as follows. Due to the small carrier density relative to the flat bands, as we dope away from the correlated states, electrons (holes) in the Dirac band only populate the zero-energy Dirac Landau level, contributing $C_{\text{D}}=2$ ($C_{\text{D}}=-2$) on the electron (hole) side even at small $B$. Since the total observed Chern number $C= C_{\text{f}} +C_{\text{D}}$, where $C_{\text{f}}$ is the Chern number of the flat band states, the observations are consistent with $C_{\text{f}}=0$. FIG. \ref{fig1}c and FIG. \ref{fig1}d show $R_{xx}$ in zero magnetic field and the Hall density $n_H$ as a function of $\nu$. The superconducting state appears near half-filling 
($|\nu|=2$) on both the hole and electron sides. From the Hall density, we see that on the hole side the superconducting state is bounded by a van Hove singularity  (vHS) (indicated by a divergence in $n_H$) and a correlated state ($n_H$ exhibits `reset' behavior). In contrast, on the electron side, the superconductor is flanked by a pair of correlated states. The results agree with previous studies~\cite{hao2021electric,park2022robust}. FIG. \ref{fig1}e and FIG. \ref{fig1}f show $dV_{xx}/dI$ as a function of DC current $I_{dc}$ and magnetic field $B$ when $\nu = -2.37$ (e) and $\nu = 2.74$ (f). The critical magnetic field of hole-side superconductivity can be as high as 400~mT, and the critical current can be around 800~nA (channel width is 1 $\mu$m), both of which highlight the robustness of the superconductivity in our device. Superconductivity on the electron side is much weaker than on the hole side, with a 30 mT critical magnetic field and 100 nA critical current. No Fraunhofer interference pattern is observed in either hole or electron sides. This may be because the device is sufficiently homogeneous that it lacks superconducting percolation paths separated by normal regions, the usual origin of such interference.

\section*{Double-dome superconductivity}
In order to understand the microscopic origin of superconductivity in MATTG, it is essential to explore the superconducting phase diagram. Since our device exhibits much stronger superconductivity on the hole side than on the electron side, we primarily focus on the former in the main text. FIG. \ref{fig2}a shows the temperature dependence of $R_{xx}$. The optimal doping (i.e., the strongest superconductivity) occurs near $\nu = -2.3$ with a critical temperature around 2.5~K, which is notably high relative to other reported MATTG devices. Superconductivity is strongly suppressed near $\nu^* = -2.6$ and exhibits a double dome feature not directly observed in previous experiments. The left dome is between the filling $-3.2<\nu<\nu^*$, and the right dome is in the region $\nu^*<\nu<-2$. FIG. \ref{fig2}b shows line cuts of $R_{xx}$ when $\nu = -2.2$ and $\nu = -2.8$. There are several differences in the temperature dependence of $R_{xx}$ in the left and right domes. The normal state resistance of the right dome follows a linear temperature behavior, as shown by the dashed line, whereas the normal state resistance in the left dome deviates from linear-in-$T$ behaviour. Another difference is that the transition region from the normal to the superconducting state is sharper in the right dome than that in the left. (The temperature dependence of $R_{xx}$ at other moiré fillings are shown in Fig. \ref{sub2}.) FIG. \ref{fig2}c shows $R_{xx}$ in the $\nu-B$ plane. Like the temperature dependence of $R_{xx}$, the magnetic field dependence also indicates that superconductivity is strongly suppressed near $\nu^* = -2.6$. The maximum $T_c$ of the left dome is smaller than that of the right dome; however, the maximum critical magnetic field in the left dome can reach as high as $400$~mT, exceeding that of the right dome. Besides suppressing superconductivity near $\nu^*$, the magnetic field also induces a correlated state at $\nu = -2$. However, at zero magnetic fields, the $T\to0$ state for $\nu = -2$ is superconducting.  The connection between correlated states and superconductivity~\cite{stepanov2020untying,saito2020independent} remains a matter of significant debate. Our results demonstrate that the correlated and superconducting states are independent and indeed may compete. Fig. \ref{fig2}d displays three line cuts of $R_{xx}$ at $B = 0$ T, $0.1$ T, and $0.2$ T. It is evident that a small magnetic field induces two resistance peaks. One corresponds to the correlated state at $\nu = -2$, while another appears near $\nu^*$, marking the boundary between two superconducting regimes. The resistance peak near $\nu^*$ may indicate the emergence of a competing order near this filling, suppressing superconductivity and thereby splitting a single large superconducting dome into two parts. Alternatively, it could be attributed to an intrinsic change in the nature of the superconducting order in two distinct domes. Though weak, but we have observed double-dome like feature on the electron side superconducting phase diagram (Fig. \ref{sub8}). We have also witnessed weak double-dome superconductivity in another near magic angle TTG devices (alternating twist angle $\theta=1.51^{\circ}$,Fig. \ref{sub9})). 

\section*{Bias current dependence and $I-V$ curve hysteresis}
We also perform DC current-dependent measurements to gain additional insight into the superconducting phase diagram in MATTG. FIG. \ref{fig3}a-c shows $R_{xx}$ as a function of DC current $I_{dc}$ and moiré filling $\nu$ at three different temperatures. As shown in FIG. \ref{fig3}a, the results can be categorized into three regions. The left dome exhibits a pronounced critical current of approximately $500$ nA. The right dome also displays a strong critical current feature; however, the critical current initially increases and then decreases with increasing carrier density, a contrast that suggests that the right dome functions independently of the left. 
Finally, for the central region near moiré filling $\nu^* = -2.6$ while the resistance is zero for low $I_{dc}$, the switching to the normal state is not sharp. Moreover, as we increase the temperature from $T = 0.3$~K to $T = 1.8$~K in FIG. \ref{fig3}c, it becomes apparent that the superconducting phase on the left side nearly disappears, while the critical current decreases slightly on the right side. In the central regime, the resistance at low $I_{dc}$ is no longer zero, but the same kink-like feature (shown by the dashed line in FIG. \ref{fig3}a,b,c) at high $I_{dc}$ persists. One possibility is that the feature in the central regime is predominantly influenced by the band structure rather than the superconductivity, as it exhibits little temperature dependence. Crucially, we also find that the right dome exhibits hysteresis in the $I-V$ curve when $I_{dc}$ is swept forwards versus backwards, revealing the switching and the retrapping currents respectively, whereas the left dome and central regime show no such hysteresis. FIG. \ref{fig3}d illustrates the DC voltage difference when sweeping $I_{dc}$ in the two directions and reveals hysteresis in the range $-2.4<\nu<-2.1$, which lies within the right dome. This hysteresis can be as large as 60~nA, as depicted in Fig. \ref{fig3}f. FIG. \ref{fig3}e shows the $I-V$ curve at $\nu=-2.84$, within the left dome. The switching and the retrapping currents are identical, showing no hysteresis. Fig. \ref{sub5} displays the voltage difference on sweeping $I_{dc}$ forwards and backwards in the left dome, confirming the absence of $I-V$ hysteresis throughout.

Hysteresis in the $I-V$ curves of a superconductor can arise for several different reasons. One common explanation invokes inhomogeneity within the superconducting sample, which may give rise to a network of underdamped Josephson junctions~\cite{tinkham2004introduction}. However, we believe that this is not the case with our device: the absence of Fraunhofer-like interference patterns, high switching currents, and large critical magnetic fields all indicate that the device is clean and homogeneous. Furthermore, the systematic dependence of hysteresis on doping and the contrasting behaviour in the two domes also suggests that it has an intrinsic origin. An alternative explanation for $I-V$ hysteresis can be ``self heating''~\cite{tinkham2003hysteretic,gurevich1987self,courtois2008origin}. In this context, one possible explanation is that there are more quasi-particles involved in conducting heat within the left dome than in the right dome. This helps facilitate the equilibration of electron and bath temperatures when sweeping $I_{dc}$ downwards from the normal state. An important clue in this regard is provided by scanning tunneling microscopy (STM) measurements, which report a transition from a `U' - shaped tunneling gap to a `V' - shaped tunneling gap~\cite{kim2022evidence}. Note that STM experiments are single-gated, and hence trace a diagonal line in the $\nu-D$ plane, which should be kept in mind when comparing results. The presence of a `V' - shaped tunneling gap close to $\nu=-3$ suggests that low-energy quasi-particles could enhance thermal conductivity in this doping, consequently reducing self-heating and suppressing $I-V$ hysteresis as we observe in the left dome at non-zero $D$. In contrast, the `U'-shaped behaviour for doping close to $\nu =-2$ is characteristic of a fully-gapped superconductor with a much lower density of thermally-conducting quasiparticles and hence poor equilibration. This can lead to hysteretic $I-V$ behaviour, consistent with our experimental findings in the right dome.

\begin{figure*}
  \centering
  \includegraphics[width= 1\textwidth]{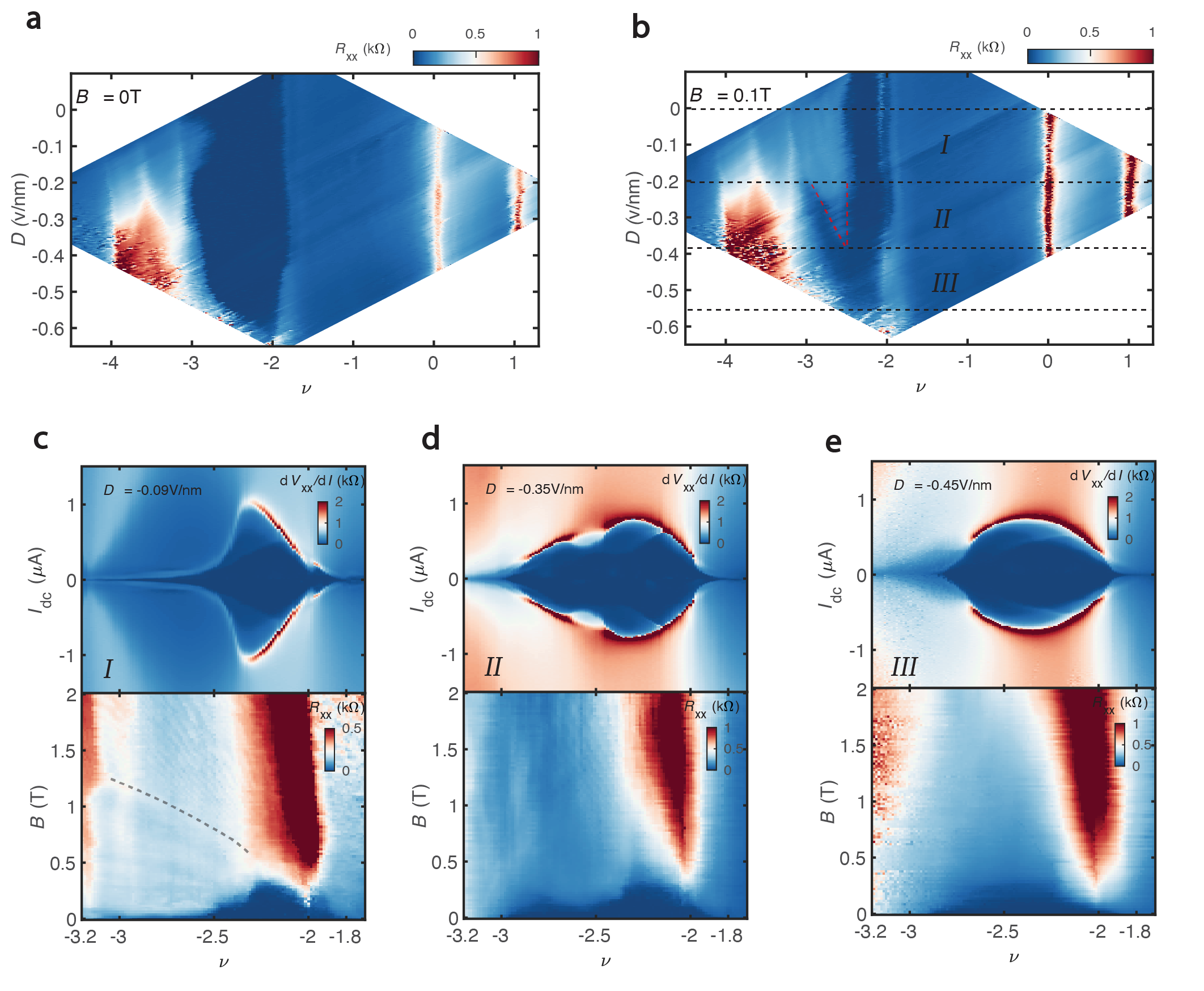}
  \caption{\textbf{Displacement field tunable double dome superconductivity ($T$ = 100 mK).} 
  \textbf{a,b}, $\nu-D$ mapping at $B=0$T (\textbf{a}) and $B=0.1$T (\textbf{b}). \textbf{c,d,e}, Differential resistance (top panel) versus $\nu$ and $I_{dc}$ at different displacement field $D=-0.09V$/{nm} (\textbf{a}), $D=-0.35V$/nm (\textbf{b}), $D=-0.45V$/nm (\textbf{c}) and corresponding Landau fan (bottom panel). In a low displacement region (\textbf{c}), the Dirac band, isolated alongside a flat band, generates Dirac Landau levels, complementing Landau levels stemming from the flat band. Charge transfer between the two sets of Landau levels gives rise to the curved feature in the lower left portion of the $\nu-B$ plane (marked with a dashed line as a guide to the eye.) In this region, a single superconducting dome exists, and the critical current of the superconducting state sharply decreases with increasing doping. The Dirac band is already hybridized with flat bands in the high displacement field regime (\textbf{e}). Landau levels solely originate from this hybridized band. In addition, only one superconducting dome exists, and the superconducting region shrinks as the displacement field increases until the superconducting dome vanishes entirely. The Dirac band is hybridizing with a flat band in the intermediate displacement field region (\textbf{d}) and the double dome superconductivity appears.
}
\label{fig4}
\end{figure*}

\begin{figure*}
  \centering
  \includegraphics[width=1\textwidth]{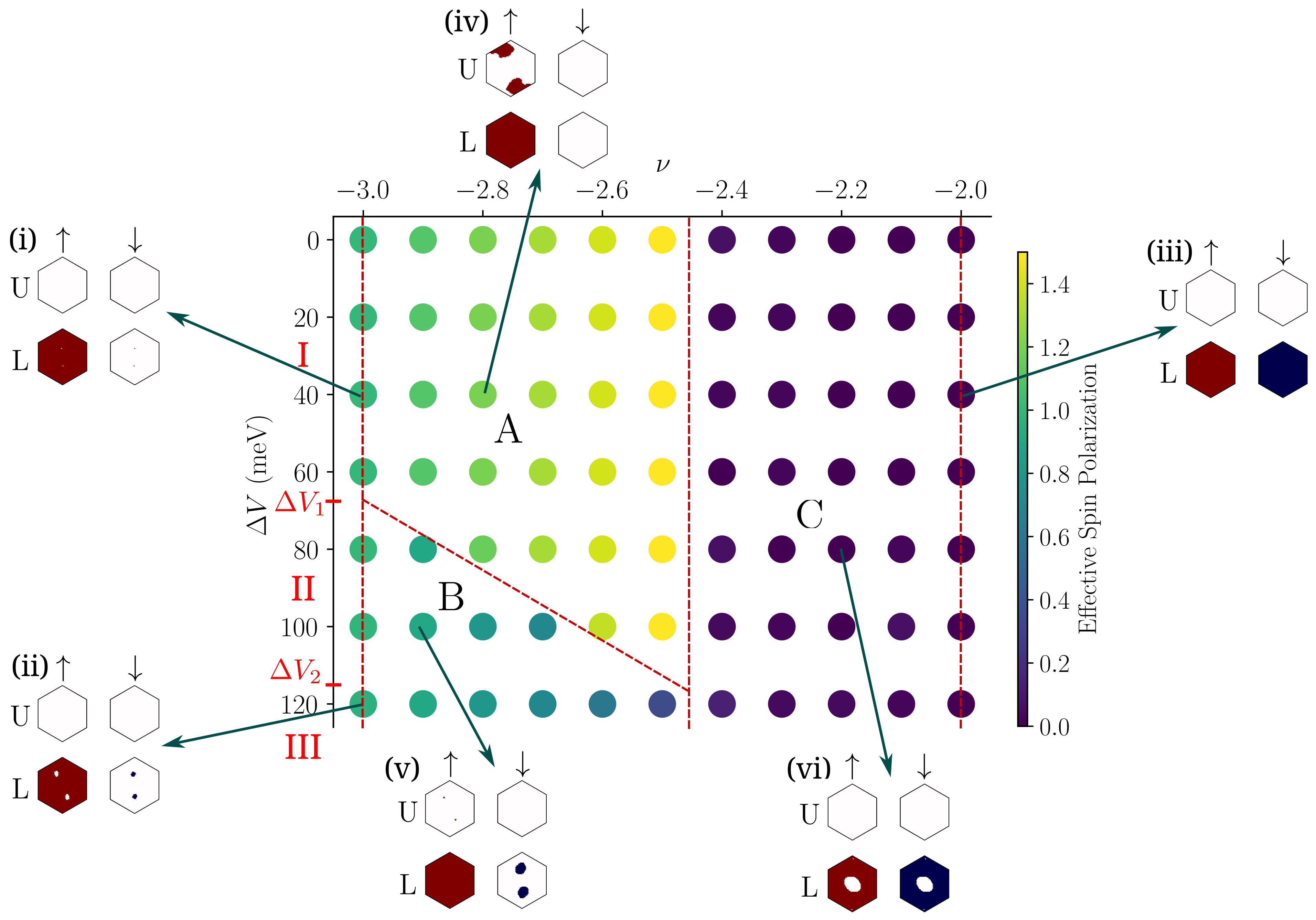}
  \caption{\textbf{Hartree-Fock phase diagram and Fermi surface  reconstruction across $\nu^*$}. The main figure shows the effective spin polarization of the Hartree-Fock ground state as a function of the filling $\nu$ and interlayer potential $\Delta V$. The latter serves a proxy for the displacement field $D$; regimes of $\Delta V$ consistent with regions I-III of Fig.~\ref{fig4} are indicated. The phase diagram shows a characteristic wedge (labeled $A$) where the effective spin polarization is peaked, which coincides with the region of suppressed superconductivity seen experimentally in FIG \ref{fig4}.b.  Insets (i)-(vi) show representative examples of the normal state Fermi surfaces (FSs) in six distinct regimes. The FSs occur in the upper (U) and lower (L) IKS bands and are labelled by their effective spin ($\uparrow/\downarrow$). \textbf{(i)} and \textbf{(ii)} show the FS for $\nu=-3$ and low and high $D$, respectively. The presence of a fully filled lower IKS band in a single effective spin sector for low $D$ and charge transfer to the lower IKS band in the opposite effective spin sector for high $D$ are consistent with `reset' behaviour in $R_{xx}$ only for low $D$ in FIG~\ref{fig1}d. \textbf{(iii)}~shows the FS for $\nu=-2$, where the lower IKS band is fully filled in both effective spin sectors. \textbf{(iv)}~in wedge $A$ doped electrons  enter the upper IKS band of the same effective spin sector that is (mostly) filled at $\nu=-3$. \textbf{(v)}~For $-3< \nu \lesssim \nu^*$ and at high displacement fields (the region labelled $B$), doped electrons form FSs in the lower IKS band of the opposite effective spin sector. This doping regime coincides with the left superconducting dome. \textbf{(vi)}~for $\nu^*\lesssim \nu <-2$ and all $D$ (region $C$) we find a pair of identical FSs in the lower IKS bands of both effective spin sectors, consistent with a twofold Landau fan degeneracy. This doping regime  corresponds to the right superconducting dome. The change in the FS structure across $\nu^*$ modifies the possibilities for superconducting pairing: whereas the FS in region $B$ only allows pairing within a single effective spin sector, the FS in region $C$ allows pairing between either the same or different effective spin sectors. Both regions $B$ and $C$ involve only the lower IKS band. The coincidence of suppressed superconductivity with the wedge $A$ of doping into the upper IKS band suggests that pairing is less robust here. Together, these  results provide a scenario for the origin of double dome superconductivity with distinct pairing channels in each dome.
  }
  \label{figure5}
\end{figure*}

\section*{Displacement field tunability of double-dome behavior}

In contrast to MATBG, MATTG exhibits a Dirac band coexisting with flat bands, offering a band structure that is adjustable through the application of a displacement field~\cite{park2021tunable,hao2021electric,shen2023dirac}. In FIG. \ref{fig4}, we study how displacement fields affect the double-dome superconductivity. FIG. \ref{fig4}a shows $R_{xx}$ as a function of displacement field $D$ and moiré filling $\nu$ in zero magnetic field. As the dark blue region shows, a superconducting region exists on the hole side. The displacement field can be divided into three regions according to the behavior of the superconducting dome. In Region I, the displacement field is low and the Dirac and flat bands are not totally hybridized. Landau levels originating from the Dirac band are visible as indicated by the curved feature in FIG. \ref{fig4}c. (The curvature is linked to varying charge transfer between Dirac and flat band Landau levels at fixed total density. Indeed, Hartree-Fock calculations show charge transfer cascades between the Dirac band and flat band sectors in MATTG \cite{wang2023kekule}.) In this region, there is only one superconducting dome, shown in FIG. \ref{fig4}c. Critical current and critical magnetic field sharply decrease as carrier density increases, starting from $\nu =-2.35$.

Region II is at intermediate displacement fields, when the Dirac band is strongly hybridized with the flat band, indicated by the absence of 
discernible signatures of Dirac band Landau levels as shown in FIG. \ref{fig4}d. In this region, superconductivity exhibits a double-dome feature in both $I_{dc}$ and magnetic field dependence measurements. Region III occurs at higher displacement fields, where the Dirac and flat band hybridization is even stronger, and the Dirac-band Landau level features are again absent, as shown by FIG. \ref{fig4}e. 
In this region, the system again displays a single superconducting dome (FIG. \ref{fig4}c) that shrinks with increasing displacement field before finally disappearing. However, the characteristics of this single dome are sharply distinct from those observed in the first region. The critical current gradually increases and then slowly decreases with an increase in carrier density instead of sharply decreasing. 

FIG. \ref{fig4}b shows the $R_{xx}$ in the $\nu-D$ plane when $B=0.1$~T. The results can again be divided into three regions. When $\lvert D \lvert<D_1$, the superconductivity of the left part is already destroyed, while the right part is still superconducting. The results agree with FIG. \ref{fig4}c where superconductivity in the left part is very weak. When $D_1< \lvert D \lvert <D_2$, two superconducting domes are bridged by a resistive region near $\nu^*$. When $D_2< \lvert D \lvert <D_3$, only one superconducting dome is observed and it shrinks with increasing displacement field, before finally disappearing when $\lvert D \lvert>D_3$. (More displacement field data is provided in FIG. \ref{sub6}).

We can get two key pieces of information from the displacement field dependence measurement: (1) the Dirac band  gradually hybridizes with the flat band as $D$ is increased; (2) Double-dome superconductivity appears in conjunction with the disappearance of Dirac band Landau levels.

\section*{Hartree-Fock studies of normal-state properties across $\nu^*$} 

In order to gain theoretical insight into the experimental observation of double-dome superconductivity, we turn to Hartree-Fock calculations on MATTG. As can be seen in FIG. \ref{fig2}b, the temperature dependence of $R_{xx}$ at fillings $\nu=-2.2$ and $\nu=-2.8$ shows clear differences far above the superconducting critical temperature itself. This suggests that the normal parent state of the superconductor undergoes changes between these fillings, and across $\nu^*$. This can be captured by computing the Hartree-Fock ground state, and hence the Fermi surface from which the superconductor emerges. 

In the presence of the experimentally realistic amounts of strain necessary to capture the Landau fan degeneracy at $\nu=-2$ and the semimetallic behaviour at $\nu=0$,  our calculations (see Methods) indicate the normal state 
of MATTG to be an incommensurate Kekulé spiral (IKS)~\cite{Kwan2021} at all non-zero integer fillings $\nu$~\cite{wang2023kekule}. IKS order, which is also ubiquitous in strained MATBG~\cite{Kwan2021}, involves a spiral in the graphene valley pseudospin degrees of freedom, and spontaneously breaks valley symmetry by the formation of spatially-modulated inter-valley coherence (IVC). This leads to a multiscale order consisting of a Kekulé charge pattern on the graphene scale (due to IVC) that rotates on the moiré scale with a wavelength incommensurate with the superlattice period (due to the spiral order). IKS order has  been directly observed in STM experiments in MATBG~\cite{Nuckolls2023} and MATTG~\cite{Kim2023}.

FIG. \ref{figure5} (main figure) shows the effective spin polarization of the Hartree-Fock ground state in the range of fillings $\nu=-3$ to $\nu=-2$ for varying interlayer potential $\Delta V$, which serves as a proxy for the displacement field $D$. The effective spin polarization corresponds to either the physical spin or the valley-filtered spin polarization, depending on whether the intervalley Hund's coupling is ferromagnetic or antiferromagnetic~\cite{Kwan2021}. 
Note that since the IKS bands involve IVC, they cannot be given a global valley label. The Hartree-Fock ground state exhibits IKS order in the entire region of the phase diagram. In addition, we find a clear wedge where the effective spin polarization is peaked between $\nu=-3$ and $\nu\approx\nu^*$. This wedge shrinks as the displacement field is applied, mirroring the wedge of suppressed superconductivity seen in FIG. \ref{fig4}b. 

Schematically, as shown in FIG. \ref{figure5} (insets), the Hartree-Fock band structure below charge neutrality consists of an upper and lower IKS band in each of the effective spin sectors. At $\nu=-3$ and small $\Delta V$, the Hartree-Fock state consists of a single filled lower IKS  band in one effective spin sector, that we will label ``up'' for specificity (FIG. \ref{figure5}i). At larger $\Delta V$ there is some charge transfer between the two lower IKS bands, leading to a small Fermi surface at $\nu=-3$ (FIG. \ref{figure5}ii). This is consistent with experimental observation of `reset' behaviour at $\nu=-3$ only at small displacement fields ~\cite{park2021tunable,hao2021electric}.

Upon doping from $\nu=-3$ towards neutrality, there are two possibilities: (1) The upper IKS band in the ``up" effective spin sector starts filling (FIG. \ref{figure5}iv), or (2) the lower IKS  band in the ``down" effective spin sector starts filling (FIG. \ref{figure5}v). Whether option (1) or (2) occurs depends on the balance of kinetic versus exchange interaction energy. For small $\Delta V$ (region A), the small bandwidth means option (1) is energetically favourable, while for large $\Delta V$ (region B), the larger bandwidth favours option (2). The regime where option (1) is favoured tracks the wedge feature. For $\nu>\nu^*$, the lower IKS bands in both effective spin sectors are equally occupied (FIG. \ref{figure5}vi). 

The entire range of the phase diagram shown in FIG. \ref{figure5} (main figure) exhibits superconductivity as seen in FIG. \ref{fig4}a. However,  comparison with FIG. \ref{fig4}b shows that the wedge where the theory predicts maximal effective spin polarization (region A) has the lowest critical magnetic field (the suppression of superconductivity also coincides with a dip in the critical temperature as seen in FIG. \ref{fig2}a). The most robust superconductivity is seen for $\nu^*\lesssim\nu<-2$ (region C). 

The Hartree-Fock state for $-3<\nu\lesssim\nu^*$ only has a single partially filled IKS band (hence a single effective spin flavor), while the state for $\nu^*\lesssim\nu<-2$ has two IKS Fermi surfaces (hence two effective spins flavors). Fermionic antisymmetry dictates that the superconducting order parameter for $-3<\nu\lesssim\nu^*$ be odd parity. On the other hand, for $\nu^*\lesssim\nu<-2$ we have two effective spin flavors and the order parameter can have either even or odd parity. STM measurements support nodal superconductivity for $-3<\nu\lesssim\nu^*$ and nodeless superconductivity for $\nu^*\lesssim\nu<-2$~\cite{kim2022evidence}. The measurements in FIG. \ref{fig2}a show a higher critical temperature in the right dome compared to the left dome. This is consistent with our theory: The possible order parameters for the single band case in the left dome are necessarily intra-effective spin. On the other hand, for the two-band case in the right dome we may have either inter- or intra-effective spin pairing. Hence, the pairing symmetries for the left dome (region B) are a  subset of those in the right dome (region C). The higher critical temperature for the right dome as well as the distinct STM signatures of the two domes suggest that inter-effective spin pairing is  realized in the right dome and is more robust than intra-effective spin pairing.  In both these cases, the superconductivity involves the lower IKS bands. The region of suppressed superconductivity coincides with the wedge in which we find the the Fermi surface to lie in an upper IKS band (region A), suggesting that this is antagonistic to the formation of strong superconducting order.

\section*{Discussion and conclusion}
The most striking result we report is the direct observation of double-dome superconductivity in hole-doped MATTG ($-3\lesssim\nu\lesssim-2$) under a displacement field. This adds a new entry to the list of unconventional phenomena observed in MATTG and expands the roster of superconductors known to show a double-dome feature as a function of doping. Through extensive electrical transport measurements we further distinguish several features of the two adjacent superconducting domes in MATTG. Notably, the strongly hysteretic $I-V$ behaviour and relative robustness of superconductivity in the right dome ($\nu^*<\nu \lesssim -2$)  contrast with the lack of hysteresis and relatively weak superconductivity in the left dome ($-3\lesssim \nu<\nu^*$). An appealing explanation for this is that there is a paucity of low-energy quasiparticles that can drive thermal equilibration in the right as compared to the left, which can in turn be rationalized in terms of a change in the nature of superconductivity from nodeless to nodal as the electron density is decreased across $\nu^*$. The latter observation would be in striking accord with STM measurements on MATTG that conjectured a transition from nodeless to nodal superconductivity on the basis of the evolution of the shape of the gap function in a similar range of doping where we find the two-dome feature at finite $D$. 

We gain further insight by tracking the appearance and subsequent disappearance of the double-dome feature with increasing displacement field. Coupled with Hartree-Fock simulations of the normal state, this provides additional clues as to the nature of superconductivity in the two domes. 
Our numerical studies find robust incommensurate Kekulé spiral order proximate to $\nu=-2$, again consistent with STM experiments, which exhibits with a clear wedge of enhanced effective spin polarization in the normal state around $\nu^*$ and in a window of displacement fields. We also infer a change in the possible superconducting pairing states in the two domes: whereas the right dome admits pairing between electrons with the same and opposite effective spins, only the former is possible in the left dome. In light of this and our  hysteresis data it is tempting to conjecture that MATTG hosts two types of pairing, a relatively robust nodeless even-parity pairing and a weaker nodal odd-parity pairing that are associated respectively with the right and left domes. The even-parity pairing obeys a modified Anderson's theorem (see Methods), making it stable to moiré scale disorder, potentially explaining its robustness. Finally,  our theoretical results indicate that in the $\nu-D$ region corresponding to either dome, doped electrons enter lower IKS bands, whereas the region of suppressed superconductivity coincides with the wedge where they enter the upper IKS band. This suggests that the latter is less susceptible to pairing, which may help identify the pairing mechanism in MATTG. Our work thus opens a significant new chapter in the phenomenology of MATTG, and motivates future work to assemble the several tantalizingly consistent pieces of evidence into a coherent picture of its superconducting phase diagram.

\bibliography{twodome}

\begin{thebibliography}{49}%
\makeatletter
\providecommand \@ifxundefined [1]{%
 \@ifx{#1\undefined}
}%
\providecommand \@ifnum [1]{%
 \ifnum #1\expandafter \@firstoftwo
 \else \expandafter \@secondoftwo
 \fi
}%
\providecommand \@ifx [1]{%
 \ifx #1\expandafter \@firstoftwo
 \else \expandafter \@secondoftwo
 \fi
}%
\providecommand \natexlab [1]{#1}%
\providecommand \enquote  [1]{``#1''}%
\providecommand \bibnamefont  [1]{#1}%
\providecommand \bibfnamefont [1]{#1}%
\providecommand \citenamefont [1]{#1}%
\providecommand \href@noop [0]{\@secondoftwo}%
\providecommand \href [0]{\begingroup \@sanitize@url \@href}%
\providecommand \@href[1]{\@@startlink{#1}\@@href}%
\providecommand \@@href[1]{\endgroup#1\@@endlink}%
\providecommand \@sanitize@url [0]{\catcode `\\12\catcode `\$12\catcode `\&12\catcode `\#12\catcode `\^12\catcode `\_12\catcode `\%12\relax}%
\providecommand \@@startlink[1]{}%
\providecommand \@@endlink[0]{}%
\providecommand \url  [0]{\begingroup\@sanitize@url \@url }%
\providecommand \@url [1]{\endgroup\@href {#1}{\urlprefix }}%
\providecommand \urlprefix  [0]{URL }%
\providecommand \Eprint [0]{\href }%
\providecommand \doibase [0]{https://doi.org/}%
\providecommand \selectlanguage [0]{\@gobble}%
\providecommand \bibinfo  [0]{\@secondoftwo}%
\providecommand \bibfield  [0]{\@secondoftwo}%
\providecommand \translation [1]{[#1]}%
\providecommand \BibitemOpen [0]{}%
\providecommand \bibitemStop [0]{}%
\providecommand \bibitemNoStop [0]{.\EOS\space}%
\providecommand \EOS [0]{\spacefactor3000\relax}%
\providecommand \BibitemShut  [1]{\csname bibitem#1\endcsname}%
\let\auto@bib@innerbib\@empty
\bibitem [{\citenamefont {Koike}\ \emph {et~al.}(1991)\citenamefont {Koike}, \citenamefont {Watanabe}, \citenamefont {Noji} \emph {et~al.}}]{koike1991effects}%
  \BibitemOpen
  \bibfield  {author} {\bibinfo {author} {\bibfnamefont {Y.}~\bibnamefont {Koike}}, \bibinfo {author} {\bibfnamefont {N.}~\bibnamefont {Watanabe}}, \bibinfo {author} {\bibfnamefont {T.}~\bibnamefont {Noji}}, \emph {et~al.},\ }\bibfield  {title} {\bibinfo {title} {Effects of the cu-site substitution on the anomalous x dependence of {T$_c$} in {La$_{2-x}$Ba$_x$CuO$_4$}},\ }\href {https://doi.org/10.1016/0038-1098(91)90366-4} {\bibfield  {journal} {\bibinfo  {journal} {Solid State Commun.}\ }\textbf {\bibinfo {volume} {78}},\ \bibinfo {pages} {511} (\bibinfo {year} {1991})}\BibitemShut {NoStop}%
\bibitem [{\citenamefont {Grissonnanche}\ \emph {et~al.}(2014)\citenamefont {Grissonnanche}, \citenamefont {Cyr-Choini{\ifmmode\grave{e}\else\`{e}\fi}re}, \citenamefont {Lalibert{\ifmmode\acute{e}\else\'{e}\fi}} \emph {et~al.}}]{grissonnanche2014direct}%
  \BibitemOpen
  \bibfield  {author} {\bibinfo {author} {\bibfnamefont {G.}~\bibnamefont {Grissonnanche}}, \bibinfo {author} {\bibfnamefont {O.}~\bibnamefont {Cyr-Choini{\ifmmode\grave{e}\else\`{e}\fi}re}}, \bibinfo {author} {\bibfnamefont {F.}~\bibnamefont {Lalibert{\ifmmode\acute{e}\else\'{e}\fi}}}, \emph {et~al.},\ }\bibfield  {title} {\bibinfo {title} {{Direct measurement of the upper critical field in cuprate superconductors}},\ }\href {https://doi.org/10.1038/ncomms4280} {\bibfield  {journal} {\bibinfo  {journal} {Nat. Commun.}\ }\textbf {\bibinfo {volume} {5}},\ \bibinfo {pages} {1} (\bibinfo {year} {2014})}\BibitemShut {NoStop}%
\bibitem [{\citenamefont {Yuan}\ \emph {et~al.}(2003)\citenamefont {Yuan}, \citenamefont {Grosche}, \citenamefont {Deppe} \emph {et~al.}}]{yuan2003observation}%
  \BibitemOpen
  \bibfield  {author} {\bibinfo {author} {\bibfnamefont {H.~Q.}\ \bibnamefont {Yuan}}, \bibinfo {author} {\bibfnamefont {F.~M.}\ \bibnamefont {Grosche}}, \bibinfo {author} {\bibfnamefont {M.}~\bibnamefont {Deppe}}, \emph {et~al.},\ }\bibfield  {title} {\bibinfo {title} {Observation of two distinct superconducting phases in {CeCu$_2$Si$_2$}},\ }\href {https://doi.org/10.1126/science.1091648} {\bibfield  {journal} {\bibinfo  {journal} {Science}\ }\textbf {\bibinfo {volume} {302}},\ \bibinfo {pages} {2104} (\bibinfo {year} {2003})}\BibitemShut {NoStop}%
\bibitem [{\citenamefont {Grosche}\ \emph {et~al.}(2000)\citenamefont {Grosche}, \citenamefont {Agarwal}, \citenamefont {Julian} \emph {et~al.}}]{grosche2000anomalous}%
  \BibitemOpen
  \bibfield  {author} {\bibinfo {author} {\bibfnamefont {F.~M.}\ \bibnamefont {Grosche}}, \bibinfo {author} {\bibfnamefont {P.}~\bibnamefont {Agarwal}}, \bibinfo {author} {\bibfnamefont {S.~R.}\ \bibnamefont {Julian}}, \emph {et~al.},\ }\bibfield  {title} {\bibinfo {title} {Anomalous low temperature states in {CeNi$_2$Ge$_2$} and {CePd$_2$Si$_2$}},\ }\href {https://doi.org/10.1088/0953-8984/12/32/101} {\bibfield  {journal} {\bibinfo  {journal} {J. Phys.: Condens. Matter}\ }\textbf {\bibinfo {volume} {12}},\ \bibinfo {pages} {L533} (\bibinfo {year} {2000})}\BibitemShut {NoStop}%
\bibitem [{\citenamefont {Zhang}\ \emph {et~al.}(2017)\citenamefont {Zhang}, \citenamefont {Liu}, \citenamefont {Ying} \emph {et~al.}}]{zhang2017observation}%
  \BibitemOpen
  \bibfield  {author} {\bibinfo {author} {\bibfnamefont {J.}~\bibnamefont {Zhang}}, \bibinfo {author} {\bibfnamefont {F.-L.}\ \bibnamefont {Liu}}, \bibinfo {author} {\bibfnamefont {T.-P.}\ \bibnamefont {Ying}}, \emph {et~al.},\ }\bibfield  {title} {\bibinfo {title} {{Observation of two superconducting domes under pressure in tetragonal FeS}},\ }\href {https://doi.org/10.1038/s41535-017-0050-7} {\bibfield  {journal} {\bibinfo  {journal} {npj Quantum Mater.}\ }\textbf {\bibinfo {volume} {2}},\ \bibinfo {pages} {1} (\bibinfo {year} {2017})}\BibitemShut {NoStop}%
\bibitem [{\citenamefont {Zhang}\ \emph {et~al.}(2021)\citenamefont {Zhang}, \citenamefont {Chen}, \citenamefont {Zhou} \emph {et~al.}}]{zhang2021pressure}%
  \BibitemOpen
  \bibfield  {author} {\bibinfo {author} {\bibfnamefont {Z.}~\bibnamefont {Zhang}}, \bibinfo {author} {\bibfnamefont {Z.}~\bibnamefont {Chen}}, \bibinfo {author} {\bibfnamefont {Y.}~\bibnamefont {Zhou}}, \emph {et~al.},\ }\bibfield  {title} {\bibinfo {title} {{Pressure-induced reemergence of superconductivity in the topological kagome metal $\mathrm{Cs}{\mathrm{V}}_{3}{\mathrm{Sb}}_{5}$}},\ }\href {https://doi.org/10.1103/PhysRevB.103.224513} {\bibfield  {journal} {\bibinfo  {journal} {Phys. Rev. B}\ }\textbf {\bibinfo {volume} {103}},\ \bibinfo {pages} {224513} (\bibinfo {year} {2021})}\BibitemShut {NoStop}%
\bibitem [{\citenamefont {Das}\ and\ \citenamefont {Panagopoulos}(2016)}]{das2016two}%
  \BibitemOpen
  \bibfield  {author} {\bibinfo {author} {\bibfnamefont {T.}~\bibnamefont {Das}}\ and\ \bibinfo {author} {\bibfnamefont {C.}~\bibnamefont {Panagopoulos}},\ }\bibfield  {title} {\bibinfo {title} {{Two types of superconducting domes in unconventional superconductors}},\ }\href {https://doi.org/10.1088/1367-2630/18/10/103033} {\bibfield  {journal} {\bibinfo  {journal} {New J. Phys.}\ }\textbf {\bibinfo {volume} {18}},\ \bibinfo {pages} {103033} (\bibinfo {year} {2016})}\BibitemShut {NoStop}%
\bibitem [{\citenamefont {Cao}\ \emph {et~al.}(2018{\natexlab{a}})\citenamefont {Cao}, \citenamefont {Fatemi}, \citenamefont {Demir} \emph {et~al.}}]{cao2018correlated}%
  \BibitemOpen
  \bibfield  {author} {\bibinfo {author} {\bibfnamefont {Y.}~\bibnamefont {Cao}}, \bibinfo {author} {\bibfnamefont {V.}~\bibnamefont {Fatemi}}, \bibinfo {author} {\bibfnamefont {A.}~\bibnamefont {Demir}}, \emph {et~al.},\ }\bibfield  {title} {\bibinfo {title} {{Correlated insulator behaviour at half-filling in magic-angle graphene superlattices}},\ }\href {https://doi.org/10.1038/nature26154} {\bibfield  {journal} {\bibinfo  {journal} {Nature}\ }\textbf {\bibinfo {volume} {556}},\ \bibinfo {pages} {80} (\bibinfo {year} {2018}{\natexlab{a}})}\BibitemShut {NoStop}%
\bibitem [{\citenamefont {Cao}\ \emph {et~al.}(2018{\natexlab{b}})\citenamefont {Cao}, \citenamefont {Fatemi}, \citenamefont {Fang} \emph {et~al.}}]{cao2018unconventional}%
  \BibitemOpen
  \bibfield  {author} {\bibinfo {author} {\bibfnamefont {Y.}~\bibnamefont {Cao}}, \bibinfo {author} {\bibfnamefont {V.}~\bibnamefont {Fatemi}}, \bibinfo {author} {\bibfnamefont {S.}~\bibnamefont {Fang}}, \emph {et~al.},\ }\bibfield  {title} {\bibinfo {title} {{Unconventional superconductivity in magic-angle graphene superlattices}},\ }\href {https://doi.org/10.1038/nature26160} {\bibfield  {journal} {\bibinfo  {journal} {Nature}\ }\textbf {\bibinfo {volume} {556}},\ \bibinfo {pages} {43} (\bibinfo {year} {2018}{\natexlab{b}})}\BibitemShut {NoStop}%
\bibitem [{\citenamefont {Yankowitz}\ \emph {et~al.}(2019)\citenamefont {Yankowitz}, \citenamefont {Chen}, \citenamefont {Polshyn} \emph {et~al.}}]{yankowitz2019tuning}%
  \BibitemOpen
  \bibfield  {author} {\bibinfo {author} {\bibfnamefont {M.}~\bibnamefont {Yankowitz}}, \bibinfo {author} {\bibfnamefont {S.}~\bibnamefont {Chen}}, \bibinfo {author} {\bibfnamefont {H.}~\bibnamefont {Polshyn}}, \emph {et~al.},\ }\bibfield  {title} {\bibinfo {title} {{Tuning superconductivity in twisted bilayer graphene}},\ }\href {https://doi.org/10.1126/science.aav1910} {\bibfield  {journal} {\bibinfo  {journal} {Science}\ }\textbf {\bibinfo {volume} {363}},\ \bibinfo {pages} {1059} (\bibinfo {year} {2019})}\BibitemShut {NoStop}%
\bibitem [{\citenamefont {Park}\ \emph {et~al.}(2021{\natexlab{a}})\citenamefont {Park}, \citenamefont {Cao}, \citenamefont {Watanabe} \emph {et~al.}}]{park2021tunable}%
  \BibitemOpen
  \bibfield  {author} {\bibinfo {author} {\bibfnamefont {J.~M.}\ \bibnamefont {Park}}, \bibinfo {author} {\bibfnamefont {Y.}~\bibnamefont {Cao}}, \bibinfo {author} {\bibfnamefont {K.}~\bibnamefont {Watanabe}}, \emph {et~al.},\ }\bibfield  {title} {\bibinfo {title} {{Tunable strongly coupled superconductivity in magic-angle twisted trilayer graphene}},\ }\href {https://doi.org/10.1038/s41586-021-03192-0} {\bibfield  {journal} {\bibinfo  {journal} {Nature}\ }\textbf {\bibinfo {volume} {590}},\ \bibinfo {pages} {249} (\bibinfo {year} {2021}{\natexlab{a}})}\BibitemShut {NoStop}%
\bibitem [{\citenamefont {Hao}\ \emph {et~al.}(2021)\citenamefont {Hao}, \citenamefont {Zimmerman}, \citenamefont {Ledwith} \emph {et~al.}}]{hao2021electric}%
  \BibitemOpen
  \bibfield  {author} {\bibinfo {author} {\bibfnamefont {Z.}~\bibnamefont {Hao}}, \bibinfo {author} {\bibfnamefont {A.~M.}\ \bibnamefont {Zimmerman}}, \bibinfo {author} {\bibfnamefont {P.}~\bibnamefont {Ledwith}}, \emph {et~al.},\ }\bibfield  {title} {\bibinfo {title} {{Electric field{\textendash}tunable superconductivity in alternating-twist magic-angle trilayer graphene}},\ }\href {https://doi.org/10.1126/science.abg0399} {\bibfield  {journal} {\bibinfo  {journal} {Science}\ }\textbf {\bibinfo {volume} {371}},\ \bibinfo {pages} {1133} (\bibinfo {year} {2021})}\BibitemShut {NoStop}%
\bibitem [{\citenamefont {Zhang}\ \emph {et~al.}(2022)\citenamefont {Zhang}, \citenamefont {Polski}, \citenamefont {Lewandowski} \emph {et~al.}}]{zhang2022promotion}%
  \BibitemOpen
  \bibfield  {author} {\bibinfo {author} {\bibfnamefont {Y.}~\bibnamefont {Zhang}}, \bibinfo {author} {\bibfnamefont {R.}~\bibnamefont {Polski}}, \bibinfo {author} {\bibfnamefont {C.}~\bibnamefont {Lewandowski}}, \emph {et~al.},\ }\bibfield  {title} {\bibinfo {title} {{Promotion of superconductivity in magic-angle graphene multilayers}},\ }\href {https://doi.org/10.1126/science.abn8585} {\bibfield  {journal} {\bibinfo  {journal} {Science}\ }\textbf {\bibinfo {volume} {377}},\ \bibinfo {pages} {1538} (\bibinfo {year} {2022})}\BibitemShut {NoStop}%
\bibitem [{\citenamefont {Park}\ \emph {et~al.}(2022)\citenamefont {Park}, \citenamefont {Cao}, \citenamefont {Xia} \emph {et~al.}}]{park2022robust}%
  \BibitemOpen
  \bibfield  {author} {\bibinfo {author} {\bibfnamefont {J.~M.}\ \bibnamefont {Park}}, \bibinfo {author} {\bibfnamefont {Y.}~\bibnamefont {Cao}}, \bibinfo {author} {\bibfnamefont {L.-Q.}\ \bibnamefont {Xia}}, \emph {et~al.},\ }\bibfield  {title} {\bibinfo {title} {{Robust superconductivity in magic-angle multilayer graphene family}},\ }\href {https://doi.org/10.1038/s41563-022-01287-1} {\bibfield  {journal} {\bibinfo  {journal} {Nat. Mater.}\ }\textbf {\bibinfo {volume} {21}},\ \bibinfo {pages} {877} (\bibinfo {year} {2022})}\BibitemShut {NoStop}%
\bibitem [{\citenamefont {Sharpe}\ \emph {et~al.}(2019)\citenamefont {Sharpe}, \citenamefont {Fox}, \citenamefont {Barnard} \emph {et~al.}}]{sharpe2019emergent}%
  \BibitemOpen
  \bibfield  {author} {\bibinfo {author} {\bibfnamefont {A.~L.}\ \bibnamefont {Sharpe}}, \bibinfo {author} {\bibfnamefont {E.~J.}\ \bibnamefont {Fox}}, \bibinfo {author} {\bibfnamefont {A.~W.}\ \bibnamefont {Barnard}}, \emph {et~al.},\ }\bibfield  {title} {\bibinfo {title} {{Emergent ferromagnetism near three-quarters filling in twisted bilayer graphene}},\ }\href {https://doi.org/10.1126/science.aaw3780} {\bibfield  {journal} {\bibinfo  {journal} {Science}\ }\textbf {\bibinfo {volume} {365}},\ \bibinfo {pages} {605} (\bibinfo {year} {2019})}\BibitemShut {NoStop}%
\bibitem [{\citenamefont {Serlin}\ \emph {et~al.}(2020)\citenamefont {Serlin}, \citenamefont {Tschirhart}, \citenamefont {Polshyn} \emph {et~al.}}]{serlin2020intrinsic}%
  \BibitemOpen
  \bibfield  {author} {\bibinfo {author} {\bibfnamefont {M.}~\bibnamefont {Serlin}}, \bibinfo {author} {\bibfnamefont {C.~L.}\ \bibnamefont {Tschirhart}}, \bibinfo {author} {\bibfnamefont {H.}~\bibnamefont {Polshyn}}, \emph {et~al.},\ }\bibfield  {title} {\bibinfo {title} {{Intrinsic quantized anomalous Hall effect in a moir{\ifmmode\acute{e}\else\'{e}\fi} heterostructure}},\ }\href {https://doi.org/10.1126/science.aay5533} {\bibfield  {journal} {\bibinfo  {journal} {Science}\ }\textbf {\bibinfo {volume} {367}},\ \bibinfo {pages} {900} (\bibinfo {year} {2020})}\BibitemShut {NoStop}%
\bibitem [{\citenamefont {Lin}\ \emph {et~al.}(2022)\citenamefont {Lin}, \citenamefont {Zhang}, \citenamefont {Morissette} \emph {et~al.}}]{lin2022spin}%
  \BibitemOpen
  \bibfield  {author} {\bibinfo {author} {\bibfnamefont {J.-X.}\ \bibnamefont {Lin}}, \bibinfo {author} {\bibfnamefont {Y.-H.}\ \bibnamefont {Zhang}}, \bibinfo {author} {\bibfnamefont {E.}~\bibnamefont {Morissette}}, \emph {et~al.},\ }\bibfield  {title} {\bibinfo {title} {{Spin-orbit{\textendash}driven ferromagnetism at half moir{\ifmmode\acute{e}\else\'{e}\fi} filling in magic-angle twisted bilayer graphene}},\ }\href {https://doi.org/10.1126/science.abh2889} {\bibfield  {journal} {\bibinfo  {journal} {Science}\ }\textbf {\bibinfo {volume} {375}},\ \bibinfo {pages} {437} (\bibinfo {year} {2022})}\BibitemShut {NoStop}%
\bibitem [{\citenamefont {Nuckolls}\ \emph {et~al.}(2020)\citenamefont {Nuckolls}, \citenamefont {Oh}, \citenamefont {Wong} \emph {et~al.}}]{nuckolls2020strongly}%
  \BibitemOpen
  \bibfield  {author} {\bibinfo {author} {\bibfnamefont {K.~P.}\ \bibnamefont {Nuckolls}}, \bibinfo {author} {\bibfnamefont {M.}~\bibnamefont {Oh}}, \bibinfo {author} {\bibfnamefont {D.}~\bibnamefont {Wong}}, \emph {et~al.},\ }\bibfield  {title} {\bibinfo {title} {{Strongly correlated Chern insulators in magic-angle twisted bilayer graphene}},\ }\href {https://doi.org/10.1038/s41586-020-3028-8} {\bibfield  {journal} {\bibinfo  {journal} {Nature}\ }\textbf {\bibinfo {volume} {588}},\ \bibinfo {pages} {610} (\bibinfo {year} {2020})}\BibitemShut {NoStop}%
\bibitem [{\citenamefont {Das}\ \emph {et~al.}(2021)\citenamefont {Das}, \citenamefont {Lu}, \citenamefont {Herzog-Arbeitman} \emph {et~al.}}]{das2021symmetry}%
  \BibitemOpen
  \bibfield  {author} {\bibinfo {author} {\bibfnamefont {I.}~\bibnamefont {Das}}, \bibinfo {author} {\bibfnamefont {X.}~\bibnamefont {Lu}}, \bibinfo {author} {\bibfnamefont {J.}~\bibnamefont {Herzog-Arbeitman}}, \emph {et~al.},\ }\bibfield  {title} {\bibinfo {title} {{Symmetry-broken Chern insulators and Rashba-like Landau-level crossings in magic-angle bilayer graphene}},\ }\href {https://doi.org/10.1038/s41567-021-01186-3} {\bibfield  {journal} {\bibinfo  {journal} {Nat. Phys.}\ }\textbf {\bibinfo {volume} {17}},\ \bibinfo {pages} {710} (\bibinfo {year} {2021})}\BibitemShut {NoStop}%
\bibitem [{\citenamefont {Saito}\ \emph {et~al.}(2021)\citenamefont {Saito}, \citenamefont {Ge}, \citenamefont {Rademaker} \emph {et~al.}}]{saito2021hofstadter}%
  \BibitemOpen
  \bibfield  {author} {\bibinfo {author} {\bibfnamefont {Y.}~\bibnamefont {Saito}}, \bibinfo {author} {\bibfnamefont {J.}~\bibnamefont {Ge}}, \bibinfo {author} {\bibfnamefont {L.}~\bibnamefont {Rademaker}}, \emph {et~al.},\ }\bibfield  {title} {\bibinfo {title} {{Hofstadter subband ferromagnetism and symmetry-broken Chern insulators in twisted bilayer graphene}},\ }\href {https://doi.org/10.1038/s41567-020-01129-4} {\bibfield  {journal} {\bibinfo  {journal} {Nat. Phys.}\ }\textbf {\bibinfo {volume} {17}},\ \bibinfo {pages} {478} (\bibinfo {year} {2021})}\BibitemShut {NoStop}%
\bibitem [{\citenamefont {Wu}\ \emph {et~al.}(2021)\citenamefont {Wu}, \citenamefont {Zhang}, \citenamefont {Watanabe} \emph {et~al.}}]{wu2021chern}%
  \BibitemOpen
  \bibfield  {author} {\bibinfo {author} {\bibfnamefont {S.}~\bibnamefont {Wu}}, \bibinfo {author} {\bibfnamefont {Z.}~\bibnamefont {Zhang}}, \bibinfo {author} {\bibfnamefont {K.}~\bibnamefont {Watanabe}}, \emph {et~al.},\ }\bibfield  {title} {\bibinfo {title} {{Chern insulators, van Hove singularities and topological flat bands in magic-angle twisted bilayer graphene}},\ }\href {https://doi.org/10.1038/s41563-020-00911-2} {\bibfield  {journal} {\bibinfo  {journal} {Nat. Mater.}\ }\textbf {\bibinfo {volume} {20}},\ \bibinfo {pages} {488} (\bibinfo {year} {2021})}\BibitemShut {NoStop}%
\bibitem [{\citenamefont {Park}\ \emph {et~al.}(2021{\natexlab{b}})\citenamefont {Park}, \citenamefont {Cao}, \citenamefont {Watanabe} \emph {et~al.}}]{park2021flavour}%
  \BibitemOpen
  \bibfield  {author} {\bibinfo {author} {\bibfnamefont {J.~M.}\ \bibnamefont {Park}}, \bibinfo {author} {\bibfnamefont {Y.}~\bibnamefont {Cao}}, \bibinfo {author} {\bibfnamefont {K.}~\bibnamefont {Watanabe}}, \emph {et~al.},\ }\bibfield  {title} {\bibinfo {title} {{Flavour Hund{'}s coupling, Chern gaps and charge diffusivity in moir{\ifmmode\acute{e}\else\'{e}\fi} graphene}},\ }\href {https://doi.org/10.1038/s41586-021-03366-w} {\bibfield  {journal} {\bibinfo  {journal} {Nature}\ }\textbf {\bibinfo {volume} {592}},\ \bibinfo {pages} {43} (\bibinfo {year} {2021}{\natexlab{b}})}\BibitemShut {NoStop}%
\bibitem [{\citenamefont {Choi}\ \emph {et~al.}(2021)\citenamefont {Choi}, \citenamefont {Kim}, \citenamefont {Peng} \emph {et~al.}}]{choi2021correlation}%
  \BibitemOpen
  \bibfield  {author} {\bibinfo {author} {\bibfnamefont {Y.}~\bibnamefont {Choi}}, \bibinfo {author} {\bibfnamefont {H.}~\bibnamefont {Kim}}, \bibinfo {author} {\bibfnamefont {Y.}~\bibnamefont {Peng}}, \emph {et~al.},\ }\bibfield  {title} {\bibinfo {title} {{Correlation-driven topological phases in magic-angle twisted bilayer graphene}},\ }\href {https://doi.org/10.1038/s41586-020-03159-7} {\bibfield  {journal} {\bibinfo  {journal} {Nature}\ }\textbf {\bibinfo {volume} {589}},\ \bibinfo {pages} {536} (\bibinfo {year} {2021})}\BibitemShut {NoStop}%
\bibitem [{\citenamefont {Pierce}\ \emph {et~al.}(2021)\citenamefont {Pierce}, \citenamefont {Xie}, \citenamefont {Park} \emph {et~al.}}]{pierce2021unconventional}%
  \BibitemOpen
  \bibfield  {author} {\bibinfo {author} {\bibfnamefont {A.~T.}\ \bibnamefont {Pierce}}, \bibinfo {author} {\bibfnamefont {Y.}~\bibnamefont {Xie}}, \bibinfo {author} {\bibfnamefont {J.~M.}\ \bibnamefont {Park}}, \emph {et~al.},\ }\bibfield  {title} {\bibinfo {title} {{Unconventional sequence of correlated Chern insulators in magic-angle twisted bilayer graphene}},\ }\href {https://doi.org/10.1038/s41567-021-01347-4} {\bibfield  {journal} {\bibinfo  {journal} {Nat. Phys.}\ }\textbf {\bibinfo {volume} {17}},\ \bibinfo {pages} {1210} (\bibinfo {year} {2021})}\BibitemShut {NoStop}%
\bibitem [{\citenamefont {Xie}\ \emph {et~al.}(2021)\citenamefont {Xie}, \citenamefont {Pierce}, \citenamefont {Park} \emph {et~al.}}]{xie2021fractional}%
  \BibitemOpen
  \bibfield  {author} {\bibinfo {author} {\bibfnamefont {Y.}~\bibnamefont {Xie}}, \bibinfo {author} {\bibfnamefont {A.~T.}\ \bibnamefont {Pierce}}, \bibinfo {author} {\bibfnamefont {J.~M.}\ \bibnamefont {Park}}, \emph {et~al.},\ }\bibfield  {title} {\bibinfo {title} {{Fractional Chern insulators in magic-angle twisted bilayer graphene}},\ }\href {https://doi.org/10.1038/s41586-021-04002-3} {\bibfield  {journal} {\bibinfo  {journal} {Nature}\ }\textbf {\bibinfo {volume} {600}},\ \bibinfo {pages} {439} (\bibinfo {year} {2021})}\BibitemShut {NoStop}%
\bibitem [{\citenamefont {Cao}\ \emph {et~al.}(2021{\natexlab{a}})\citenamefont {Cao}, \citenamefont {Rodan-Legrain}, \citenamefont {Park} \emph {et~al.}}]{cao2021nematicity}%
  \BibitemOpen
  \bibfield  {author} {\bibinfo {author} {\bibfnamefont {Y.}~\bibnamefont {Cao}}, \bibinfo {author} {\bibfnamefont {D.}~\bibnamefont {Rodan-Legrain}}, \bibinfo {author} {\bibfnamefont {J.~M.}\ \bibnamefont {Park}}, \emph {et~al.},\ }\bibfield  {title} {\bibinfo {title} {{Nematicity and competing orders in superconducting magic-angle graphene}},\ }\href {https://doi.org/10.1126/science.abc2836} {\bibfield  {journal} {\bibinfo  {journal} {Science}\ }\textbf {\bibinfo {volume} {372}},\ \bibinfo {pages} {264} (\bibinfo {year} {2021}{\natexlab{a}})}\BibitemShut {NoStop}%
\bibitem [{\citenamefont {Cao}\ \emph {et~al.}(2021{\natexlab{b}})\citenamefont {Cao}, \citenamefont {Park}, \citenamefont {Watanabe} \emph {et~al.}}]{cao2021pauli}%
  \BibitemOpen
  \bibfield  {author} {\bibinfo {author} {\bibfnamefont {Y.}~\bibnamefont {Cao}}, \bibinfo {author} {\bibfnamefont {J.~M.}\ \bibnamefont {Park}}, \bibinfo {author} {\bibfnamefont {K.}~\bibnamefont {Watanabe}}, \emph {et~al.},\ }\bibfield  {title} {\bibinfo {title} {{Pauli-limit violation and re-entrant superconductivity in moir{\ifmmode\acute{e}\else\'{e}\fi} graphene}},\ }\href {https://doi.org/10.1038/s41586-021-03685-y} {\bibfield  {journal} {\bibinfo  {journal} {Nature}\ }\textbf {\bibinfo {volume} {595}},\ \bibinfo {pages} {526} (\bibinfo {year} {2021}{\natexlab{b}})}\BibitemShut {NoStop}%
\bibitem [{\citenamefont {Kim}\ \emph {et~al.}(2022)\citenamefont {Kim}, \citenamefont {Choi}, \citenamefont {Lewandowski} \emph {et~al.}}]{kim2022evidence}%
  \BibitemOpen
  \bibfield  {author} {\bibinfo {author} {\bibfnamefont {H.}~\bibnamefont {Kim}}, \bibinfo {author} {\bibfnamefont {Y.}~\bibnamefont {Choi}}, \bibinfo {author} {\bibfnamefont {C.}~\bibnamefont {Lewandowski}}, \emph {et~al.},\ }\bibfield  {title} {\bibinfo {title} {{Evidence for unconventional superconductivity in twisted trilayer graphene}},\ }\href {https://doi.org/10.1038/s41586-022-04715-z} {\bibfield  {journal} {\bibinfo  {journal} {Nature}\ }\textbf {\bibinfo {volume} {606}},\ \bibinfo {pages} {494} (\bibinfo {year} {2022})}\BibitemShut {NoStop}%
\bibitem [{\citenamefont {Oh}\ \emph {et~al.}(2021)\citenamefont {Oh}, \citenamefont {Nuckolls}, \citenamefont {Wong} \emph {et~al.}}]{oh2021evidence}%
  \BibitemOpen
  \bibfield  {author} {\bibinfo {author} {\bibfnamefont {M.}~\bibnamefont {Oh}}, \bibinfo {author} {\bibfnamefont {K.~P.}\ \bibnamefont {Nuckolls}}, \bibinfo {author} {\bibfnamefont {D.}~\bibnamefont {Wong}}, \emph {et~al.},\ }\bibfield  {title} {\bibinfo {title} {{Evidence for unconventional superconductivity in twisted bilayer graphene}},\ }\href {https://doi.org/10.1038/s41586-021-04121-x} {\bibfield  {journal} {\bibinfo  {journal} {Nature}\ }\textbf {\bibinfo {volume} {600}},\ \bibinfo {pages} {240} (\bibinfo {year} {2021})}\BibitemShut {NoStop}%
\bibitem [{\citenamefont {Khalaf}\ \emph {et~al.}(2019)\citenamefont {Khalaf}, \citenamefont {Kruchkov}, \citenamefont {Tarnopolsky} \emph {et~al.}}]{khalaf2019magic}%
  \BibitemOpen
  \bibfield  {author} {\bibinfo {author} {\bibfnamefont {E.}~\bibnamefont {Khalaf}}, \bibinfo {author} {\bibfnamefont {A.~J.}\ \bibnamefont {Kruchkov}}, \bibinfo {author} {\bibfnamefont {G.}~\bibnamefont {Tarnopolsky}}, \emph {et~al.},\ }\bibfield  {title} {\bibinfo {title} {{Magic angle hierarchy in twisted graphene multilayers}},\ }\href {https://doi.org/10.1103/PhysRevB.100.085109} {\bibfield  {journal} {\bibinfo  {journal} {Phys. Rev. B}\ }\textbf {\bibinfo {volume} {100}},\ \bibinfo {pages} {085109} (\bibinfo {year} {2019})}\BibitemShut {NoStop}%
\bibitem [{\citenamefont {Shen}\ \emph {et~al.}(2023)\citenamefont {Shen}, \citenamefont {Ledwith}, \citenamefont {Watanabe} \emph {et~al.}}]{shen2023dirac}%
  \BibitemOpen
  \bibfield  {author} {\bibinfo {author} {\bibfnamefont {C.}~\bibnamefont {Shen}}, \bibinfo {author} {\bibfnamefont {P.~J.}\ \bibnamefont {Ledwith}}, \bibinfo {author} {\bibfnamefont {K.}~\bibnamefont {Watanabe}}, \emph {et~al.},\ }\bibfield  {title} {\bibinfo {title} {{Dirac spectroscopy of strongly correlated phases in twisted trilayer graphene}},\ }\href {https://doi.org/10.1038/s41563-022-01428-6} {\bibfield  {journal} {\bibinfo  {journal} {Nat. Mater.}\ }\textbf {\bibinfo {volume} {22}},\ \bibinfo {pages} {316} (\bibinfo {year} {2023})}\BibitemShut {NoStop}%
\bibitem [{\citenamefont {Stepanov}\ \emph {et~al.}(2020)\citenamefont {Stepanov}, \citenamefont {Das}, \citenamefont {Lu} \emph {et~al.}}]{stepanov2020untying}%
  \BibitemOpen
  \bibfield  {author} {\bibinfo {author} {\bibfnamefont {P.}~\bibnamefont {Stepanov}}, \bibinfo {author} {\bibfnamefont {I.}~\bibnamefont {Das}}, \bibinfo {author} {\bibfnamefont {X.}~\bibnamefont {Lu}}, \emph {et~al.},\ }\bibfield  {title} {\bibinfo {title} {{Untying the insulating and superconducting orders in magic-angle graphene}},\ }\href {https://doi.org/10.1038/s41586-020-2459-6} {\bibfield  {journal} {\bibinfo  {journal} {Nature}\ }\textbf {\bibinfo {volume} {583}},\ \bibinfo {pages} {375} (\bibinfo {year} {2020})}\BibitemShut {NoStop}%
\bibitem [{\citenamefont {Saito}\ \emph {et~al.}(2020)\citenamefont {Saito}, \citenamefont {Ge}, \citenamefont {Watanabe} \emph {et~al.}}]{saito2020independent}%
  \BibitemOpen
  \bibfield  {author} {\bibinfo {author} {\bibfnamefont {Y.}~\bibnamefont {Saito}}, \bibinfo {author} {\bibfnamefont {J.}~\bibnamefont {Ge}}, \bibinfo {author} {\bibfnamefont {K.}~\bibnamefont {Watanabe}}, \emph {et~al.},\ }\bibfield  {title} {\bibinfo {title} {{Independent superconductors and correlated insulators in twisted bilayer graphene}},\ }\href {https://doi.org/10.1038/s41567-020-0928-3} {\bibfield  {journal} {\bibinfo  {journal} {Nat. Phys.}\ }\textbf {\bibinfo {volume} {16}},\ \bibinfo {pages} {926} (\bibinfo {year} {2020})}\BibitemShut {NoStop}%
\bibitem [{\citenamefont {Tinkham}(2004)}]{tinkham2004introduction}%
  \BibitemOpen
  \bibfield  {author} {\bibinfo {author} {\bibfnamefont {M.}~\bibnamefont {Tinkham}},\ }\href@noop {} {\emph {\bibinfo {title} {Introduction to superconductivity}}}\ (\bibinfo  {publisher} {Courier Corporation},\ \bibinfo {year} {2004})\BibitemShut {NoStop}%
\bibitem [{\citenamefont {Tinkham}\ \emph {et~al.}(2003)\citenamefont {Tinkham}, \citenamefont {Free}, \citenamefont {Lau} \emph {et~al.}}]{tinkham2003hysteretic}%
  \BibitemOpen
  \bibfield  {author} {\bibinfo {author} {\bibfnamefont {M.}~\bibnamefont {Tinkham}}, \bibinfo {author} {\bibfnamefont {J.~U.}\ \bibnamefont {Free}}, \bibinfo {author} {\bibfnamefont {C.~N.}\ \bibnamefont {Lau}}, \emph {et~al.},\ }\bibfield  {title} {\bibinfo {title} {{Hysteretic $I\ensuremath{-}V$ curves of superconducting nanowires}},\ }\href {https://doi.org/10.1103/PhysRevB.68.134515} {\bibfield  {journal} {\bibinfo  {journal} {Phys. Rev. B}\ }\textbf {\bibinfo {volume} {68}},\ \bibinfo {pages} {134515} (\bibinfo {year} {2003})}\BibitemShut {NoStop}%
\bibitem [{\citenamefont {Gurevich}\ and\ \citenamefont {Mints}(1987)}]{gurevich1987self}%
  \BibitemOpen
  \bibfield  {author} {\bibinfo {author} {\bibfnamefont {A.~{\relax Vl}.}\ \bibnamefont {Gurevich}}\ and\ \bibinfo {author} {\bibfnamefont {R.~G.}\ \bibnamefont {Mints}},\ }\bibfield  {title} {\bibinfo {title} {{Self-heating in normal metals and superconductors}},\ }\href {https://doi.org/10.1103/RevModPhys.59.941} {\bibfield  {journal} {\bibinfo  {journal} {Rev. Mod. Phys.}\ }\textbf {\bibinfo {volume} {59}},\ \bibinfo {pages} {941} (\bibinfo {year} {1987})}\BibitemShut {NoStop}%
\bibitem [{\citenamefont {Courtois}\ \emph {et~al.}(2008)\citenamefont {Courtois}, \citenamefont {Meschke}, \citenamefont {Peltonen} \emph {et~al.}}]{courtois2008origin}%
  \BibitemOpen
  \bibfield  {author} {\bibinfo {author} {\bibfnamefont {H.}~\bibnamefont {Courtois}}, \bibinfo {author} {\bibfnamefont {M.}~\bibnamefont {Meschke}}, \bibinfo {author} {\bibfnamefont {J.~T.}\ \bibnamefont {Peltonen}}, \emph {et~al.},\ }\bibfield  {title} {\bibinfo {title} {{Origin of Hysteresis in a Proximity Josephson Junction}},\ }\href {https://doi.org/10.1103/PhysRevLett.101.067002} {\bibfield  {journal} {\bibinfo  {journal} {Phys. Rev. Lett.}\ }\textbf {\bibinfo {volume} {101}},\ \bibinfo {pages} {067002} (\bibinfo {year} {2008})}\BibitemShut {NoStop}%
\bibitem [{\citenamefont {Wang}\ \emph {et~al.}(2023)\citenamefont {Wang}, \citenamefont {Kwan}, \citenamefont {Wagner} \emph {et~al.}}]{wang2023kekule}%
  \BibitemOpen
  \bibfield  {author} {\bibinfo {author} {\bibfnamefont {Z.}~\bibnamefont {Wang}}, \bibinfo {author} {\bibfnamefont {Y.~H.}\ \bibnamefont {Kwan}}, \bibinfo {author} {\bibfnamefont {G.}~\bibnamefont {Wagner}}, \emph {et~al.},\ }\href@noop {} {\bibinfo {title} {Kekul\'e spirals and charge transfer cascades in twisted symmetric trilayer graphene}} (\bibinfo {year} {2023}),\ \Eprint {https://arxiv.org/abs/2310.16094} {arXiv:2310.16094 [cond-mat.str-el]} \BibitemShut {NoStop}%
\bibitem [{\citenamefont {Kwan}\ \emph {et~al.}(2021)\citenamefont {Kwan}, \citenamefont {Wagner}, \citenamefont {Soejima} \emph {et~al.}}]{Kwan2021}%
  \BibitemOpen
  \bibfield  {author} {\bibinfo {author} {\bibfnamefont {Y.~H.}\ \bibnamefont {Kwan}}, \bibinfo {author} {\bibfnamefont {G.}~\bibnamefont {Wagner}}, \bibinfo {author} {\bibfnamefont {T.}~\bibnamefont {Soejima}}, \emph {et~al.},\ }\bibfield  {title} {\bibinfo {title} {Kekul\'e spiral order at all nonzero integer fillings in twisted bilayer graphene},\ }\href {https://doi.org/10.1103/PhysRevX.11.041063} {\bibfield  {journal} {\bibinfo  {journal} {Phys. Rev. X}\ }\textbf {\bibinfo {volume} {11}},\ \bibinfo {pages} {041063} (\bibinfo {year} {2021})}\BibitemShut {NoStop}%
\bibitem [{\citenamefont {Nuckolls}\ \emph {et~al.}(2023)\citenamefont {Nuckolls}, \citenamefont {Lee}, \citenamefont {Oh} \emph {et~al.}}]{Nuckolls2023}%
  \BibitemOpen
  \bibfield  {author} {\bibinfo {author} {\bibfnamefont {K.~P.}\ \bibnamefont {Nuckolls}}, \bibinfo {author} {\bibfnamefont {R.~L.}\ \bibnamefont {Lee}}, \bibinfo {author} {\bibfnamefont {M.}~\bibnamefont {Oh}}, \emph {et~al.},\ }\bibfield  {title} {\bibinfo {title} {{Quantum textures of the many-body wavefunctions in magic-angle graphene}},\ }\href {https://doi.org/10.1038/s41586-023-06226-x} {\bibfield  {journal} {\bibinfo  {journal} {Nature}\ }\textbf {\bibinfo {volume} {620}},\ \bibinfo {pages} {525} (\bibinfo {year} {2023})}\BibitemShut {NoStop}%
\bibitem [{\citenamefont {Kim}\ \emph {et~al.}(2023)\citenamefont {Kim}, \citenamefont {Choi}, \citenamefont {Lantagne-Hurtubise} \emph {et~al.}}]{Kim2023}%
  \BibitemOpen
  \bibfield  {author} {\bibinfo {author} {\bibfnamefont {H.}~\bibnamefont {Kim}}, \bibinfo {author} {\bibfnamefont {Y.}~\bibnamefont {Choi}}, \bibinfo {author} {\bibfnamefont {{\ifmmode\acute{E}\else\'{E}\fi}.}~\bibnamefont {Lantagne-Hurtubise}}, \emph {et~al.},\ }\bibfield  {title} {\bibinfo {title} {{Imaging inter-valley coherent order in magic-angle twisted trilayer graphene}},\ }\href {https://doi.org/10.1038/s41586-023-06663-8} {\bibfield  {journal} {\bibinfo  {journal} {Nature}\ }\textbf {\bibinfo {volume} {623}},\ \bibinfo {pages} {942} (\bibinfo {year} {2023})}\BibitemShut {NoStop}%
\bibitem [{\citenamefont {Bistritzer}\ and\ \citenamefont {MacDonald}(2011)}]{bistritzer2011moire}%
  \BibitemOpen
  \bibfield  {author} {\bibinfo {author} {\bibfnamefont {R.}~\bibnamefont {Bistritzer}}\ and\ \bibinfo {author} {\bibfnamefont {A.~H.}\ \bibnamefont {MacDonald}},\ }\bibfield  {title} {\bibinfo {title} {{Moir{\ifmmode\acute{e}\else\'{e}\fi} bands in twisted double-layer graphene}},\ }\href {https://doi.org/10.1073/pnas.1108174108} {\bibfield  {journal} {\bibinfo  {journal} {Proc. Natl. Acad. Sci. U.S.A.}\ }\textbf {\bibinfo {volume} {108}},\ \bibinfo {pages} {12233} (\bibinfo {year} {2011})}\BibitemShut {NoStop}%
\bibitem [{\citenamefont {Kerelsky}\ \emph {et~al.}(2019)\citenamefont {Kerelsky}, \citenamefont {McGilly}, \citenamefont {Kennes} \emph {et~al.}}]{Kerelsky2019}%
  \BibitemOpen
  \bibfield  {author} {\bibinfo {author} {\bibfnamefont {A.}~\bibnamefont {Kerelsky}}, \bibinfo {author} {\bibfnamefont {L.~J.}\ \bibnamefont {McGilly}}, \bibinfo {author} {\bibfnamefont {D.~M.}\ \bibnamefont {Kennes}}, \emph {et~al.},\ }\bibfield  {title} {\bibinfo {title} {{Maximized electron interactions at the magic angle in twisted bilayer graphene}},\ }\href {https://doi.org/10.1038/s41586-019-1431-9} {\bibfield  {journal} {\bibinfo  {journal} {Nature}\ }\textbf {\bibinfo {volume} {572}},\ \bibinfo {pages} {95} (\bibinfo {year} {2019})}\BibitemShut {NoStop}%
\bibitem [{\citenamefont {Wong}\ \emph {et~al.}(2020)\citenamefont {Wong}, \citenamefont {Nuckolls}, \citenamefont {Oh} \emph {et~al.}}]{Wong2020}%
  \BibitemOpen
  \bibfield  {author} {\bibinfo {author} {\bibfnamefont {D.}~\bibnamefont {Wong}}, \bibinfo {author} {\bibfnamefont {K.~P.}\ \bibnamefont {Nuckolls}}, \bibinfo {author} {\bibfnamefont {M.}~\bibnamefont {Oh}}, \emph {et~al.},\ }\bibfield  {title} {\bibinfo {title} {{Cascade of electronic transitions in magic-angle twisted bilayer graphene}},\ }\href {https://doi.org/10.1038/s41586-020-2339-0} {\bibfield  {journal} {\bibinfo  {journal} {Nature}\ }\textbf {\bibinfo {volume} {582}},\ \bibinfo {pages} {198} (\bibinfo {year} {2020})}\BibitemShut {NoStop}%
\bibitem [{\citenamefont {Choi}\ \emph {et~al.}(2019)\citenamefont {Choi}, \citenamefont {Kemmer}, \citenamefont {Peng} \emph {et~al.}}]{Choi2019}%
  \BibitemOpen
  \bibfield  {author} {\bibinfo {author} {\bibfnamefont {Y.}~\bibnamefont {Choi}}, \bibinfo {author} {\bibfnamefont {J.}~\bibnamefont {Kemmer}}, \bibinfo {author} {\bibfnamefont {Y.}~\bibnamefont {Peng}}, \emph {et~al.},\ }\bibfield  {title} {\bibinfo {title} {{Electronic correlations in twisted bilayer graphene near the magic angle}},\ }\href {https://doi.org/10.1038/s41567-019-0606-5} {\bibfield  {journal} {\bibinfo  {journal} {Nat. Phys.}\ }\textbf {\bibinfo {volume} {15}},\ \bibinfo {pages} {1174} (\bibinfo {year} {2019})}\BibitemShut {NoStop}%
\bibitem [{\citenamefont {Xie}\ \emph {et~al.}(2019)\citenamefont {Xie}, \citenamefont {Lian}, \citenamefont {J{\ifmmode\ddot{a}\else\"{a}\fi}ck} \emph {et~al.}}]{Xie2019}%
  \BibitemOpen
  \bibfield  {author} {\bibinfo {author} {\bibfnamefont {Y.}~\bibnamefont {Xie}}, \bibinfo {author} {\bibfnamefont {B.}~\bibnamefont {Lian}}, \bibinfo {author} {\bibfnamefont {B.}~\bibnamefont {J{\ifmmode\ddot{a}\else\"{a}\fi}ck}}, \emph {et~al.},\ }\bibfield  {title} {\bibinfo {title} {{Spectroscopic signatures of many-body correlations in magic-angle twisted bilayer graphene}},\ }\href {https://doi.org/10.1038/s41586-019-1422-x} {\bibfield  {journal} {\bibinfo  {journal} {Nature}\ }\textbf {\bibinfo {volume} {572}},\ \bibinfo {pages} {101} (\bibinfo {year} {2019})}\BibitemShut {NoStop}%
\bibitem [{\citenamefont {Parker}\ \emph {et~al.}(2021)\citenamefont {Parker}, \citenamefont {Soejima}, \citenamefont {Hauschild} \emph {et~al.}}]{Parker2021}%
  \BibitemOpen
  \bibfield  {author} {\bibinfo {author} {\bibfnamefont {D.~E.}\ \bibnamefont {Parker}}, \bibinfo {author} {\bibfnamefont {T.}~\bibnamefont {Soejima}}, \bibinfo {author} {\bibfnamefont {J.}~\bibnamefont {Hauschild}}, \emph {et~al.},\ }\bibfield  {title} {\bibinfo {title} {Strain-induced quantum phase transitions in magic-angle graphene},\ }\href {https://doi.org/10.1103/PhysRevLett.127.027601} {\bibfield  {journal} {\bibinfo  {journal} {Phys. Rev. Lett.}\ }\textbf {\bibinfo {volume} {127}},\ \bibinfo {pages} {027601} (\bibinfo {year} {2021})}\BibitemShut {NoStop}%
\bibitem [{\citenamefont {Wagner}\ \emph {et~al.}(2022)\citenamefont {Wagner}, \citenamefont {Kwan}, \citenamefont {Bultinck} \emph {et~al.}}]{Wagner2022}%
  \BibitemOpen
  \bibfield  {author} {\bibinfo {author} {\bibfnamefont {G.}~\bibnamefont {Wagner}}, \bibinfo {author} {\bibfnamefont {Y.~H.}\ \bibnamefont {Kwan}}, \bibinfo {author} {\bibfnamefont {N.}~\bibnamefont {Bultinck}}, \emph {et~al.},\ }\bibfield  {title} {\bibinfo {title} {Global phase diagram of the normal state of twisted bilayer graphene},\ }\href {https://doi.org/10.1103/PhysRevLett.128.156401} {\bibfield  {journal} {\bibinfo  {journal} {Phys. Rev. Lett.}\ }\textbf {\bibinfo {volume} {128}},\ \bibinfo {pages} {156401} (\bibinfo {year} {2022})}\BibitemShut {NoStop}%
\bibitem [{\citenamefont {Anderson}(1959)}]{ANDERSON195926}%
  \BibitemOpen
  \bibfield  {author} {\bibinfo {author} {\bibfnamefont {P.}~\bibnamefont {Anderson}},\ }\bibfield  {title} {\bibinfo {title} {Theory of dirty superconductors},\ }\href {https://doi.org/https://doi.org/10.1016/0022-3697(59)90036-8} {\bibfield  {journal} {\bibinfo  {journal} {Journal of Physics and Chemistry of Solids}\ }\textbf {\bibinfo {volume} {11}},\ \bibinfo {pages} {26} (\bibinfo {year} {1959})}\BibitemShut {NoStop}%
\end{thebibliography}%

\section*{METHODS}
\subsection{Sample fabrication and measurements}

The stacks are made by using the dry-transfer technique. First, graphene and hBN are exfoliated on {SiO$_2$}/Si substrates, and monolayer graphene and proper hBN (10 - 50 nm) are identified under an optical microscope. The big monolayer graphene is precut by an AFM tip into three pieces instead of being cut by hBN to reduce strain during stacking. We used homemade poly(bisphenol A carbonate)/polydimethylsiloxane (PC/PDMS) to pick up all the flakes sequentially. First, top graphite is picked up around 100 $^\circ$C, and then top hBN is picked up around 70 $^\circ$C. We used the combined structure PDMS/PC/top graphite/top hBN to pick up the first layer of graphene at 50 $^\circ$C; afterward, the micro-stage which holds the materials chip is rotated by 1.6 - 1.65$^\circ$. Then, the second graphene piece is picked up. Subsequently, the micro-stage is rotated back to its original place, and the third graphene is picked up. Then, bottom hBN and narrow bottom graphite are picked up. The final stack is dropped down on the chip with aligned markers. The Hall bar geometry is defined by using electron-beam lithography and reactive ion etching. 5nm/60nm Cr/Au is evaporated to connect the graphene to form one-dimensional contacts.

The electronic transport measurements are performed in a dilution fridge with a base temperature of 100 mK. The temperature dependence measurement is done in a He-3 fridge. Resistance measurements are conducted using a standard lock-in technique employing a 10 nA AC current excitation at a frequency of 17 Hz. Two Yokogawa GS200 are used to apply top and bottom gate voltage to tune carrier density and displacement field. Voltage signals are taken before being amplified 100 times.

\subsection{Hall density and estimation of twist angle}
The Hall density in FIG. \ref{fig1}d is extracted from the Hall resistance $R_{xy}$. In order to reduce the effect of $R_{xx}$ on $R_{xy}$, we used $R_{xy} = (R_{xy}(B)+R_{xy}(-B))/2$. The Hall density can be extracted through the relation $R_{xy}=B/(en)$. The twist angle is linked to full moiré carrier density by $N_f=8\theta^2/\sqrt{3}a^2$, where $a= 0.246~\text{nm}$ is the lattice constant of graphene. The carrier density in the device is determined by $n=(C_{tg}V_{tg}+C_{bg}V_{bg})/e$. According to FIG. \ref{fig1}d, near $\nu=0$, the Hall density equals the carrier density, so the absolute capacitance between the bottom gate and MATTG can be extracted. The bottom gate voltage $V_{bg(full)}$ corresponding to the full filling of the flat band can be obtained from FIG. \ref{fig1}b. Full carrier density is then expressed as $N_{f}=C_{bg}V_{bg(full)}{/e}$. The twist angle of the device is $1.54^\circ$ which is the magic angle of twisted trilayer graphene.

\subsection{Extraction of $T_c$ and coherence length $\xi_{GL}$}
The critical temperature is defined as the temperature when the resistance $R_{xx}(T)$ equals to x$R_{xx}(Normal)$, x is the percentage of normal resistance $R_{xx}(Normal)$. $R_{xx}(Normal)$ is extracted by linearly fitting the high-temperature part of $R_{xx}(T)$ by $R_{xx}(Normal)=AT+B$. (The normal resistance of the left superconducting dome is not so linearly dependent on temperature, but we still use the same method to define the normal state resistance). The Ginzburg–Landau coherence length is extracted through the dependence of $T_c$ on magnetic field $B$: $T_c/T_{c0}=1-(2\pi\xi_{GL}/\Phi_0)B$. $T_{c0}$ is critical temperature when magnetic field is zero and $\Phi_0=h/2e$ is superconducting quantum flux. The Ginzburg–Landau coherence length $\xi_{GL}$ is then extracted from a linear fit of $T_c$ versus $B$. FIG. \ref{sub4} shows $\xi_{GL}$ as a function of moiré filling by using x=10\% to define $T_c$ (the corresponding error bars are evaluated by using $T_c$ defined by 6\% and 14\% of normal state resistance).

\subsection{In-plane magnetic field dependence}
The in-plane field dependence measurement is performed in an 8T-2T-2T triple-axis magnet. As the sample plane is not strictly perpendicular to the z-direction, an out-of-plane magnetic field exists when the in-plane magnetic field $B_x$ and $B_y$ are applied, and two-dimensional superconductivity is very sensitive to the out-of-plane magnetic field. So $B_z$ is needed to compensate for the out-of-plane magnetic field. The calibration of the magnetic field is done in the following way: a set of $(B_{xi}, B_{yi})$ is applied to the device, and then we tune the carrier density near the boundary of the superconducting dome and measure $V_{xx}$ as a function of $B_z$. When $V_{xx}$ is smallest, the corresponding $B_{zi}$ is our needed compensating magnetic field. A set of $(B_{xi},B_{yi},B_{zi})$ can be linearly fitted and we can express $B_z$ as $B_z = \alpha B_x + \beta B_y$ where $\alpha$ and $\beta$ are fitting parameters.

\subsection{Hartree-Fock studies}

The non-interacting band structure can be captured using a three-layer generalization of the celebrated Bistritzer-MacDonald continuum model~\cite{bistritzer2011moire}. The gate screened Coulomb interaction is incorporated via a self-consistent Hartree-Fock mean-field approach (for further details see Ref.~~\cite{wang2023kekule}). We fix the gate distance $d=25$~nm, dielectric constant $\epsilon_r=20$, and twist angle $\theta=1.56^\circ$. In addition, we include uniaxial heterostrain of strength $\epsilon=0.15\%$. This is consistent with typical values detected in many moiré graphene samples using scanning tunneling microscopy (STM)~\cite{Kerelsky2019,Wong2020,Choi2019,Xie2019}. In addition, two features of our transport data strongly suggest that our sample has strain: The absence of strong insulating behaviour at charge neutrality~\cite{Parker2021} as well as the two-fold Landau fan degeneracy at filling $\nu=-2$~\cite{Wagner2022}, which is not reproduced in realistic theoretical models of the flat bands unless strain is included. Our Hartree-Fock calculations are performed on a $10\times10$ momentum mesh, and are Hartree-Fock-interpolated to a $50 \times 50$ mesh to construct the Fermi surfaces.

\subsection{Modified time-reversal symmetry}

In region C of FIG. \ref{figure5} (main figure), including $\nu = -2$, the Hartree-Fock (HF) ground state consists of one copy of the (doped) IKS state in each of the two spin sectors (assuming a  collinear spin configuration) and as such preserves a modified time-reversal symmetry $\tilde{\mathcal{T}}=\textrm{exp}(i\Delta \phi_{\textrm{IVC}} s_z \tau_z)\mathcal{T}$, where $s_z$ and $\tau_z$ act on spin and valley degrees of freedom respectively, $\Delta \phi_{\textrm{IVC}}$ is the difference in IVC angles of the two spin sectors and $\mathcal{T}$ is the usual time-reversal operator acting on both spin and orbital degrees of freedom. More generally, the HF ground state is given by $S\ket{\psi}_{\textrm{colinear}}$ where $S$ is some valley-specific spin rotation, and the HF ground state is symmetric under  $S\textrm{exp}(i\Delta \phi_{\text{IVC}} s_z \tau_z)\mathcal{T}S^{-1}$. This modified time-reversal symmetry $\tilde{\mathcal{T}}$ leads to a generalized Anderson theorem for superconductivity \cite{ANDERSON195926}, implying robustness of the gap to disorder which preserves $\tilde{\mathcal{T}}$. Moiré scale disorder would preserve $\tilde{\mathcal{T}}$, however graphene scale disorder can break $\tilde{\mathcal{T}}$ since it couples directly to the valley $U(1)$ degree of freedom (clearly, neither form of disorder breaks the standard time-reversal symmetry $\mathcal T$). Therefore, we expect the superconductor in region C to be robust against moiré scale disorder. This is contrast to the superconductor in region B, and hence provides an appealing explanation for the differening robustness of the right and left domes.

For completeness we note that  our numerical simulations do find that in small portions of Region C, $\tilde{\mathcal{T}}$-breaking states with finite polarization  can be closely competing with IKS states; however only the IKS order is robust across most of region C.

\begin{acknowledgments}
We thank Erez Berg, Cyprian Lewandowski, Mohit Randeria, Nandini Trivedi, Jason Alicia, Oskar Vafek, and Cheng Shen for important discussions. P.K. thanks Artem Kononov for help with the cryostat. Z.Z., J.J. acknowledges funding from SNSF. M.B. acknowledges the support of SNSF Eccellenza grant No. PCEGP2\_194528, and support from the QuantERA II Programme that has received funding from the European Union’s Horizon 2020 research and innovation program under Grant Agreement No 101017733. P.K. and C.S. acknowledge support from the European Research Council (ERC) under the European Union’s Horizon 2020 research and innovation program: grant agreement No 787414, ERC-Adv TopSupra. K.W. and T.T. acknowledge support from the JSPS KAKENHI (Grant Numbers 20H00354 and 23H02052) and World Premier International Research Center Initiative (WPI), MEXT, Japan. G.W.~acknowledges funding from the University of Zurich postdoc grant FK-23-134. Z.W. acknowledges funding from Leverhulme Trust International Professorship grant (number LIP-202-014). For the purpose of Open Access, the authors have applied a CC-BY public copyright license to any Author Accepted Manuscript version arising from this submission. S.H.S. acknowledges support from EPSRC grant EP/X030881/1. S.A.P. acknowledges support from the European Research Council under the European Union Horizon 2020 Research and Innovation Programme via Grant Agreement No. 804213-TMCS, and from a Gutzwiller Fellowship at the Max Planck Institute for the Physics of Complex Systems, Dresden.

\end{acknowledgments}

\textbf{\begin{center}Author contributions\end{center}}

Z.Z. and M.B. conceived the project. J.J. fabricated the devices with Z.Z.'s help. Z.Z. has performed all the measurements and analyzed the data and M.B. supervised the project. P.K. and C.S. provided the 100 mK measurement cryostat with vector magnet. K.W. and T.T. provided the hBN crystals. Z.W. performed the numerical Hartree-Fock simulations. Z.W., G.W., S.A.P., and S.H.S. performed the theoretical analysis. Z.Z., Z.W., G.W., S.H.S., S.A.P. and M.B. wrote the manuscript with input from all authors.

\textbf{\begin{center}Data availability\end{center}}

The data supporting the findings of this study are available from the corresponding author upon reasonable request.

\onecolumngrid
\newpage

\clearpage

\section*{Supplementary Materials}

\section{Landau fans in the presence of flat bands and Dirac bands}

The Landau fans emerging from $\nu=-2$ in FIG \ref{fig4}c deviate from a linear slope. In this section we argue that this can be a consequence of charge transfer between the Dirac and flat band sectors. For Dirac fermions with a Fermi velocity $v_F$, the Landau levels in magnetic field $B$ have the form 
\begin{equation}
    E_N^\textrm{Dirac}=\pm v_F\sqrt{2|N|B},
\end{equation}
while for the flat band sector the Landau levels (LLs) are located at energies
\begin{equation}
    E_N^\textrm{flat}=\frac{B}{m^*}(N+\frac{1}{2}).
\end{equation}
We now assume the Dirac band LLs to be closely spaced compared to the flat band LLs. The density at which $Ng$ flat band LLs are filled ($g$ is the degeneracy of the flat band) is
\begin{equation}
    n=n^\textrm{flat}+n^\textrm{Dirac}=Ng\frac{N_\phi}{A}+n^\textrm{Dirac}(\mu),
\end{equation}
where $N_\phi=BA/\phi_0$ is the number of flux quanta ($A$ is the area of the sample and $\phi_0$ is the flux quantum). We have $n^\textrm{Dirac}(\mu)=\frac{\mu^2}{\pi v_F}$
and 
$\mu=E_N^\textrm{flat}$
such that
\begin{equation}
    n=\frac{Ng}{\phi_0}B+\frac{(N+1/2)^2}{\pi v_F m^{*2}}B^2,
\end{equation}
yielding a deviation from a linear slope for the Landau fans. Similarly, for the case where the flat band LLs are closely spaced compared to the Dirac band LLs, one also obtains a deviation from the linear Landau fans.

\section{Supplementary figures}

\begin{figure*}[b]
\renewcommand{\thefigure}{S1}
  \centering
  \includegraphics[width= 0.9\textwidth]{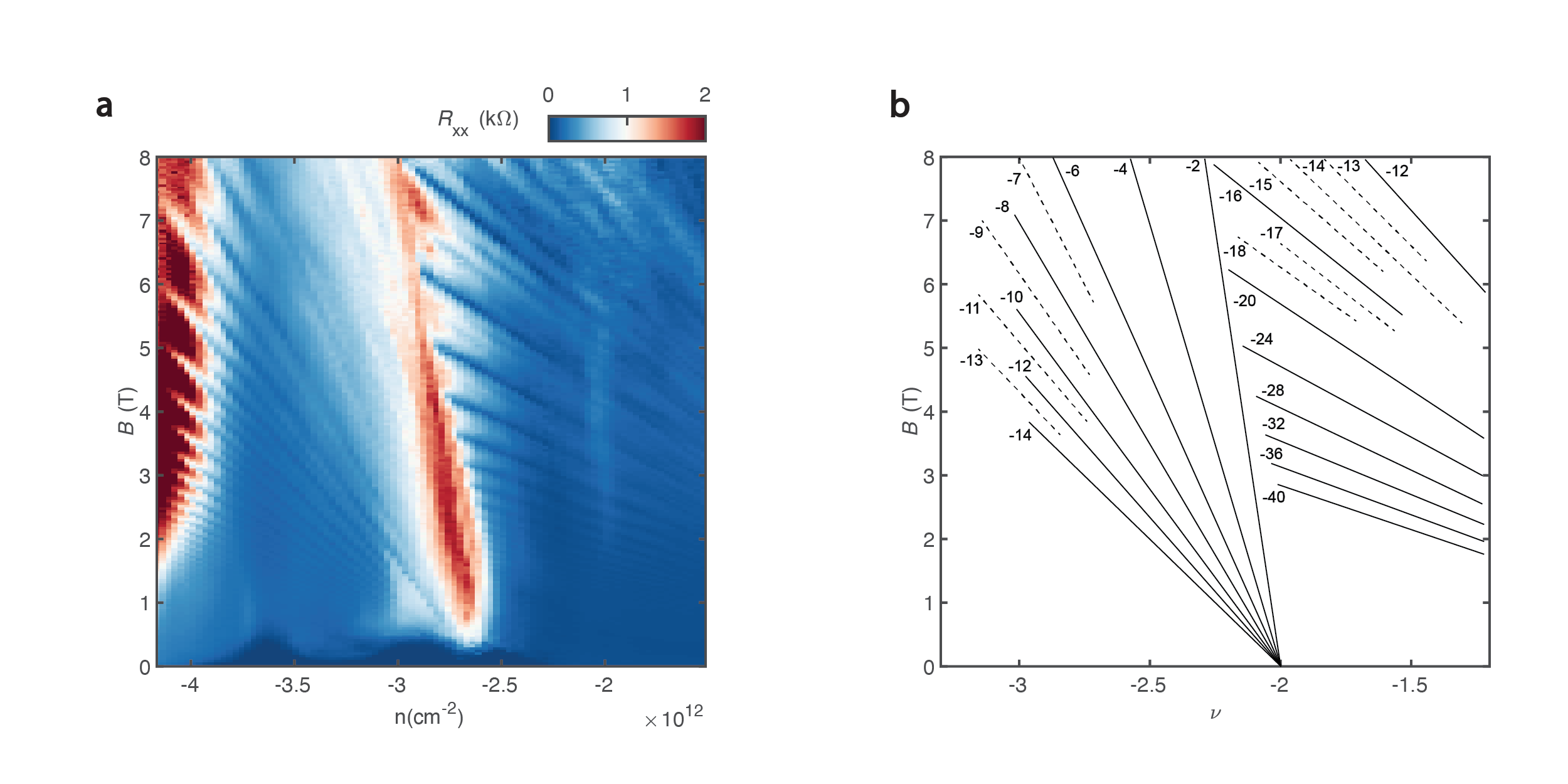}
  \caption{\textbf{Symmetry breaking at $\nu=-2$.} \textbf{a,b}, The Landau fan diagram from Fig. \ref{fig1}b in the main text zoomed in near $\nu = -2$  shows that the Landau levels stemming from $\nu = -2$ have a sequence of $-2, -4, -6$ with a twofold degeneracy. This is due to the lifting of the degeneracy of the Fermi surface by the interaction-driven symmetry breaking. The main Landau levels originating from $\nu=0$ have a four fold degeneracy without any symmetry breaking.}
  \label{sub1}
\end{figure*}

\begin{figure*}
\renewcommand{\thefigure}{S2}
  \centering
  \includegraphics[width= 0.95\textwidth]{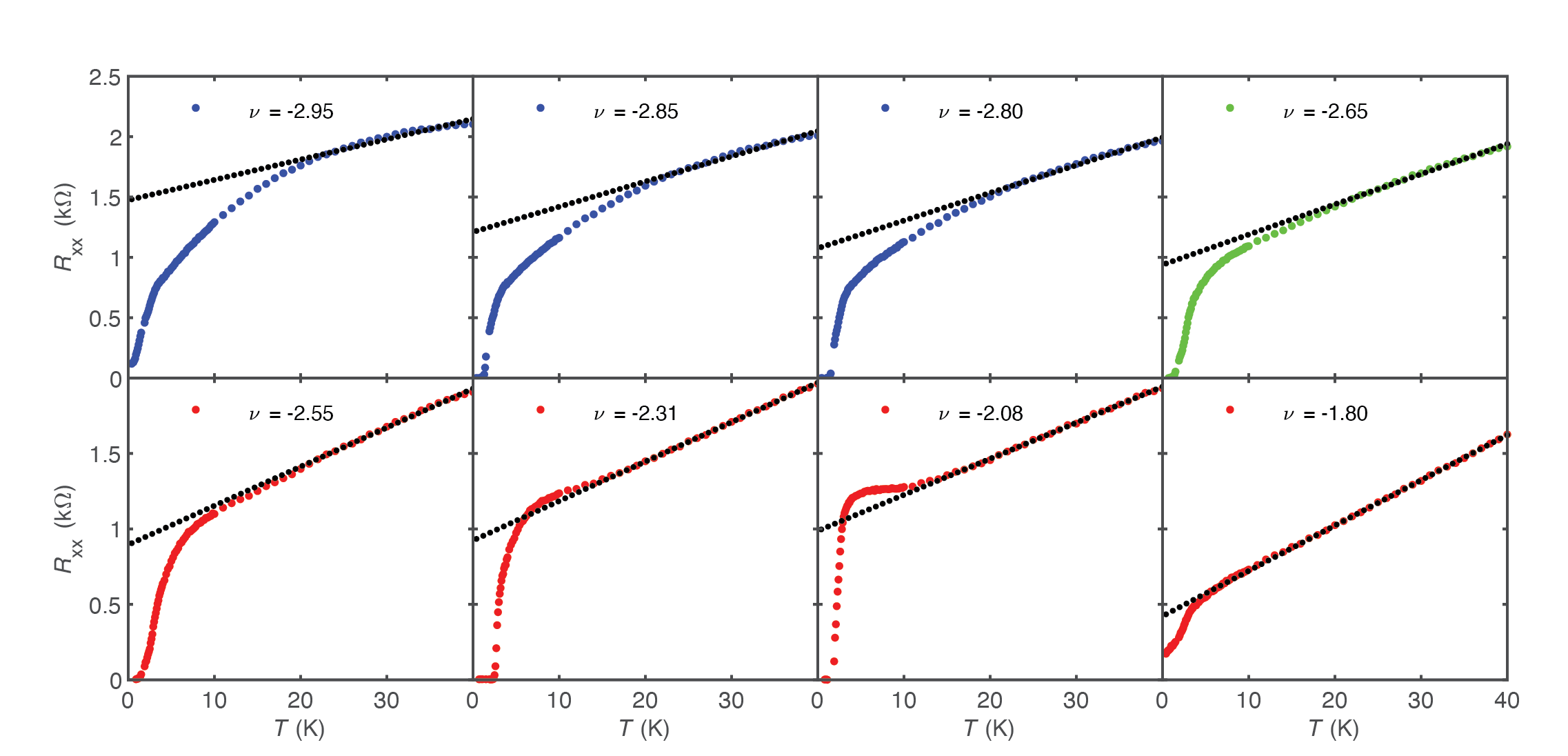}
  \caption{\textbf{Temperature dependence of $R_{xx}$.} More line cuts of $R_{xx}$ as shown in Fig.~\ref{fig2}a in the main text. The normal state of the right dome (red curves) obeys a linear temperature dependence, while that of the left dome (blue curves) deviates from linear temperature dependence. $\nu^*$ (green curve) marks the transition between the two domes.}
  \label{sub2}
\end{figure*}

\begin{figure*}
\renewcommand{\thefigure}{S3}
  \centering
  \includegraphics[width= 1\textwidth]{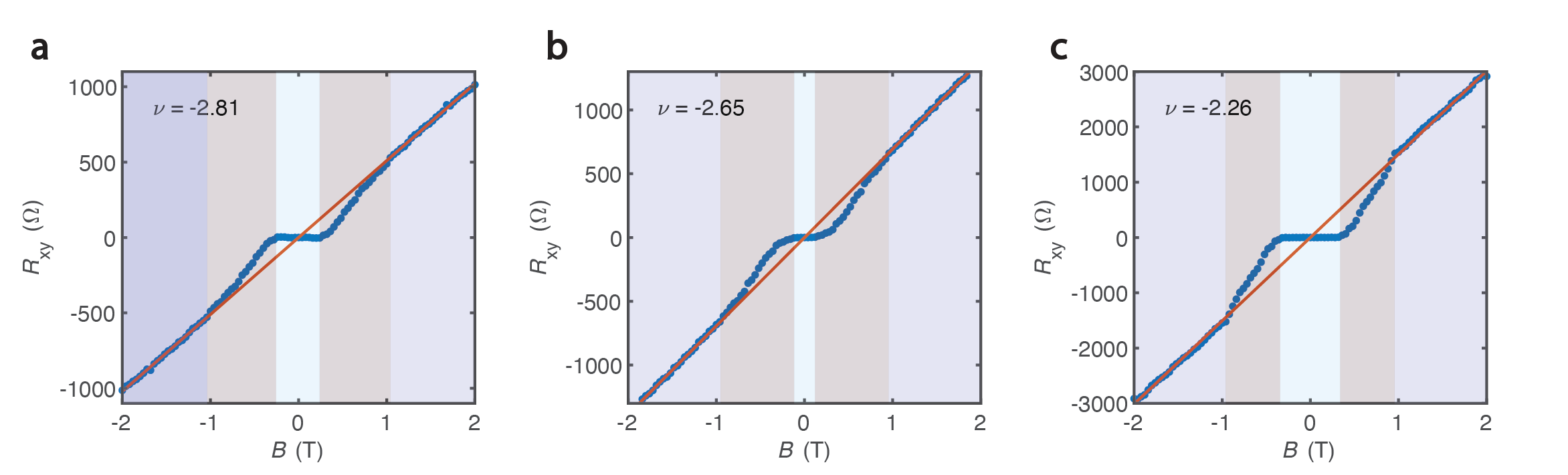}
  \caption{\textbf{Transition from superconducting state into Hall state.} $R_{xy} = (R_{xy}(B)-R_{xy}(-B))/2$ as a function of $B$ at $\nu = -2.81$ (\textbf{a}), $\nu = -2.65$ (\textbf{b}) and $\nu = -2.26$ (\textbf{c}). When the magnetic field is lower than the critical magnetic field $B_c$, the system is in the superconducting state with zero Hall resistance. When the magnetic field is high, the system transitions to a state with a linear-in-$B$ Hall resistance.} 
  \label{sub3}
\end{figure*}

\begin{figure*}
\renewcommand{\thefigure}{S4}
  \centering
  \includegraphics[width= 0.453\textwidth]{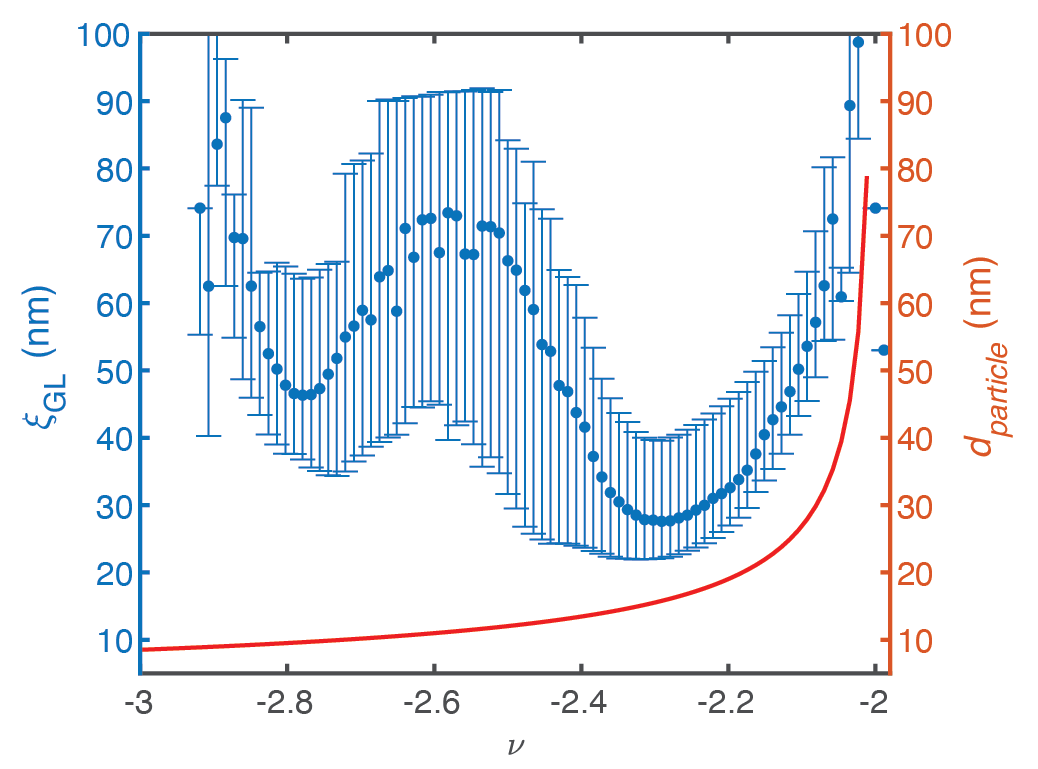}
  \caption{\textbf{Ginzburg Landau length.} Ginzburg Landau coherence length $\xi_{GL}$ as a function of $\nu$ by using $T_{c}^{10\%}$ with errorbars defined by using $T_{c}^{6\%}$ and $T_{c}^{14\%}$. The red line shows the inter-particle distance defined as $d_{particle}$ = $1/\sqrt{n^*}$ versus $\nu$. For all the density regimes, $\xi_{GL}/d_{particle}$ is between 1 and 10, which implies strong coupling superconductivity in MATTG.}
  \label{sub4}
\end{figure*}

\begin{figure*}
\renewcommand{\thefigure}{S5}
  \centering
  \includegraphics[width= 0.73\textwidth]{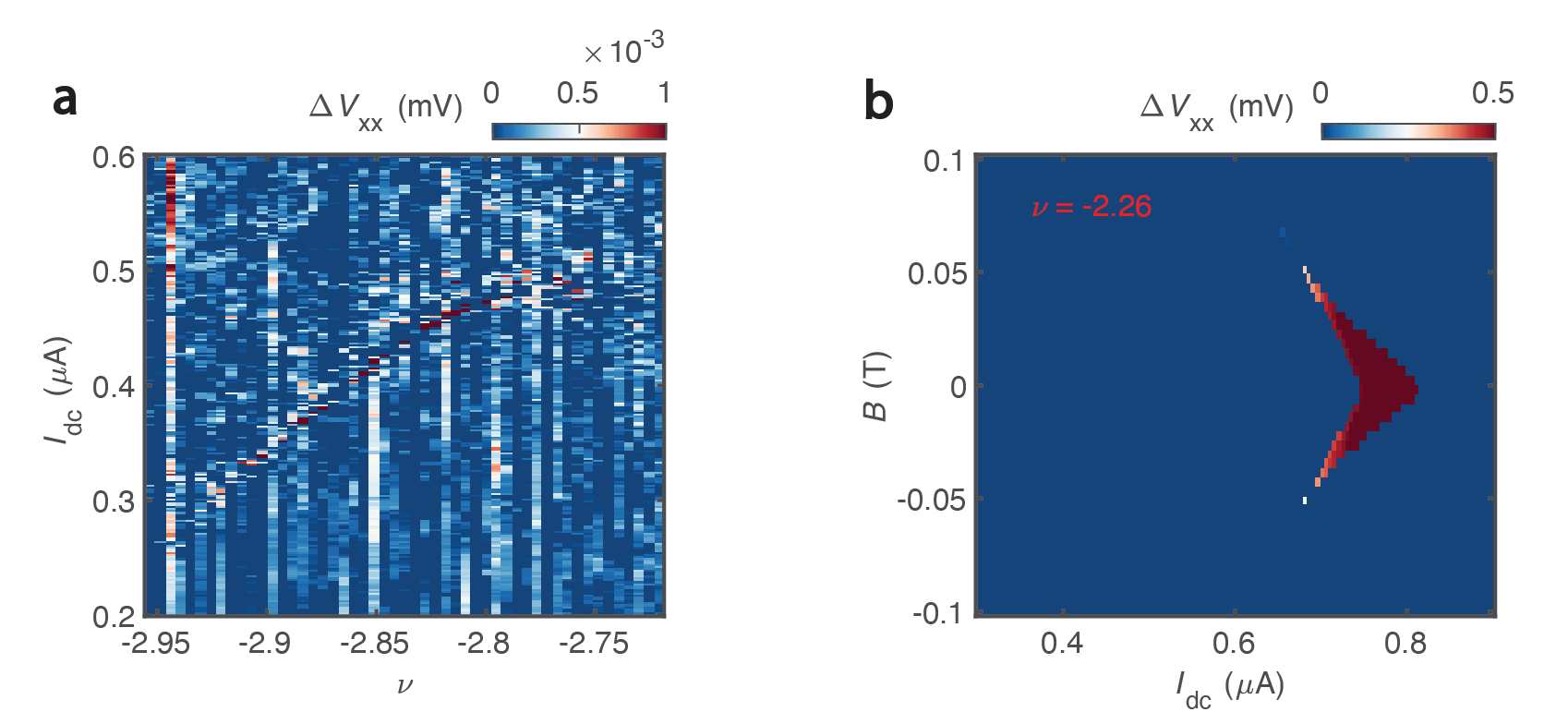}
  \caption{\textbf{Absence of $I-V$ hysteresis in the left dome and suppression of $I-V$ hysteresis in the right dome by a magnetic field.} \textbf{a}, $\Delta V_{xx}$ when sweeping $I_{dc}$ forwards and backwards in the left dome. It is clear that there is no $I - V$ hysteresis in the left dome as $\Delta V_{xx}$ is close to zero. \textbf{b}, $\Delta V_{xx}$ versus $B$ and $I_{dc}$ at $\nu = -2.26$. The hysteresis disappears as the magnetic field increases. }
  \label{sub5}
\end{figure*}

\begin{figure*}
\renewcommand{\thefigure}{S6}
  \centering
  \includegraphics[width= 0.878\textwidth]{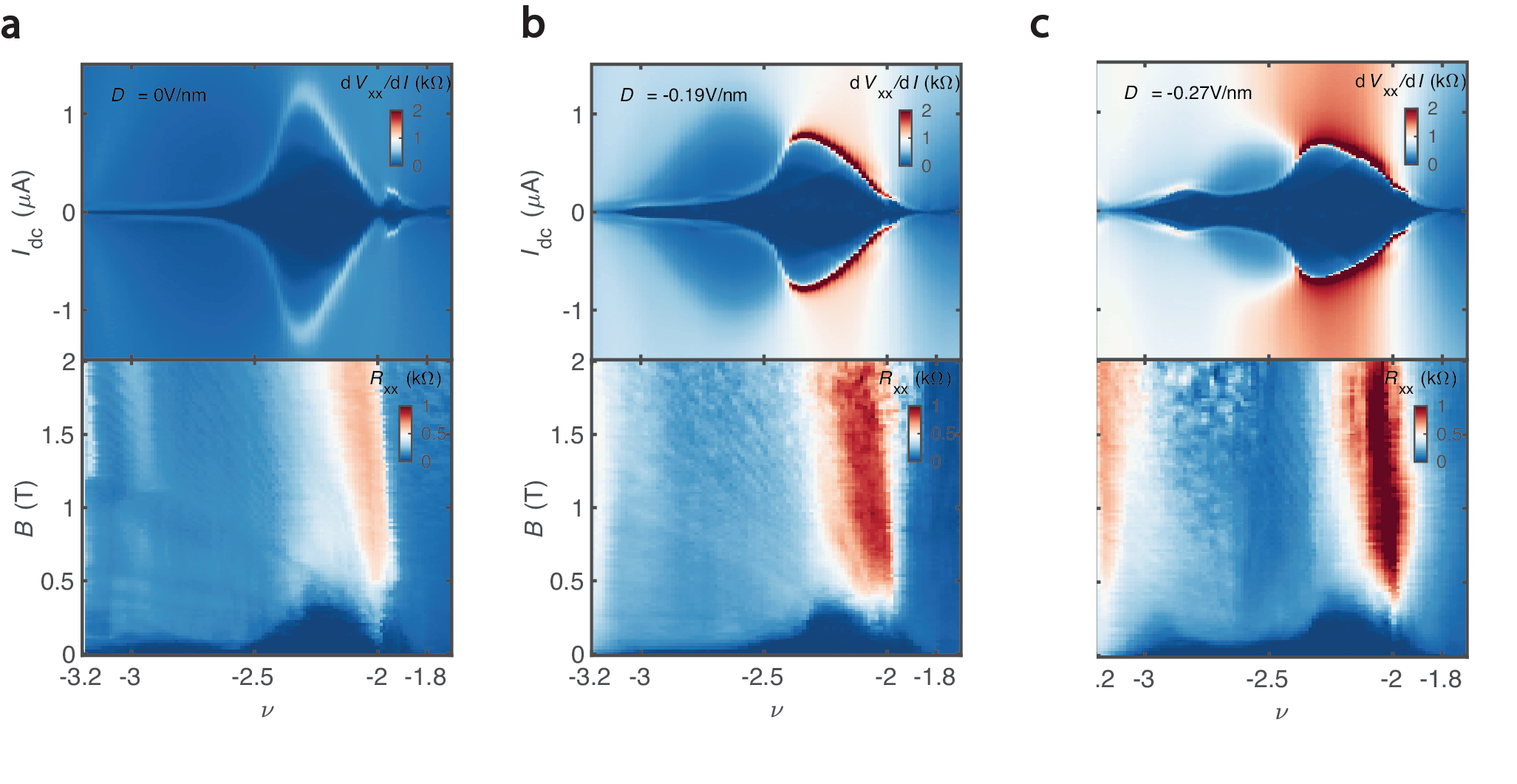}
  \caption{\textbf{Bias current and magnetic field  dependence at other displacement fields.} Differential resistance (top panel) versus $\nu$ and $I_{dc}$ at different displacement fields $D=0V$/{nm} (\textbf{a}, in region $I$), $D=-0.19V$/nm (\textbf{b}, in region $I$), $D=-0.27V$/nm (\textbf{c}, in region $II$) and corresponding Landau fan (bottom panel). Both \textbf{a} and \textbf{b} are in region $I$,  Dirac Landau levels are visible and only one superconducting dome exists with sharpely decreasing critical current for $\nu<-2.5$. \textbf{c} is in region $II$ and shows double-dome superconductivity.}
  \label{sub6}
\end{figure*}

\begin{figure*}
\renewcommand{\thefigure}{S7}
  \centering
  \includegraphics[width= 0.97\textwidth]{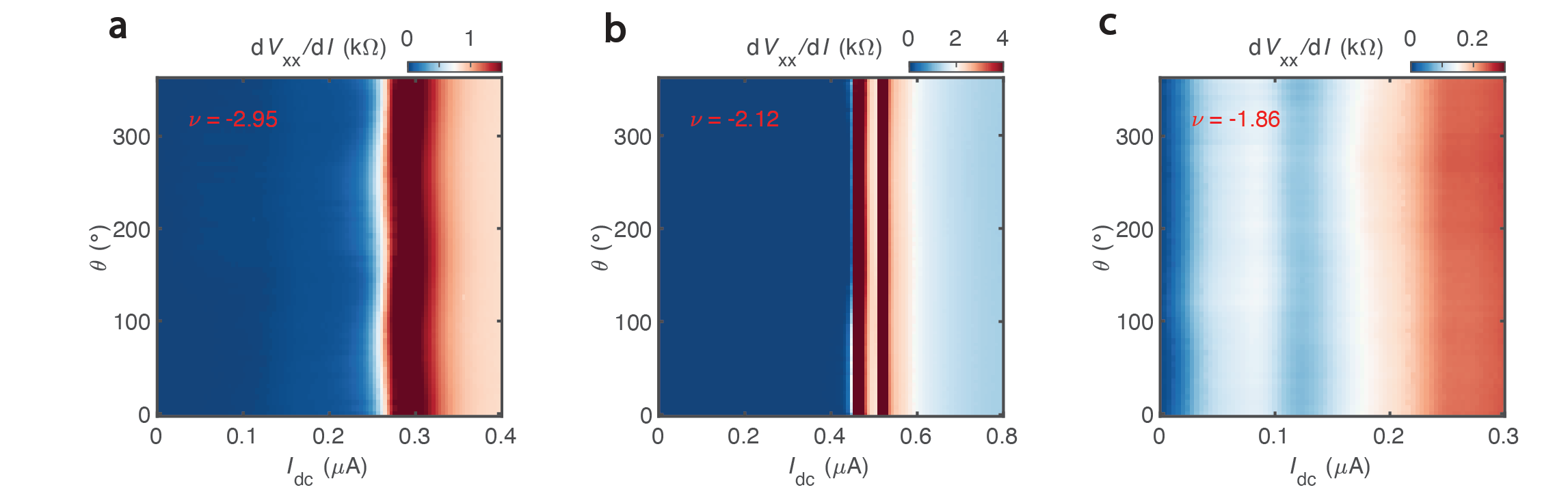}
  \caption{\textbf{In-plane magnetic field dependence measurements.} \textbf{a,b,c}, $dV_{xx}/dI$ versus the direction of the in plane magnetic field $\theta$ and $I_{dc}$ at $\nu = -2.95$ (\textbf{a}),
  $\nu = -2.12$ (\textbf{b}), $\nu = -1.86$ (\textbf{c}). The results are not dependent on the direction of $B_\parallel$. The superconductivity shows no nematicity.}
  \label{sub7}
\end{figure*}

\begin{figure*}
\renewcommand{\thefigure}{S8}
  \centering
  \includegraphics[width= 0.97\textwidth]{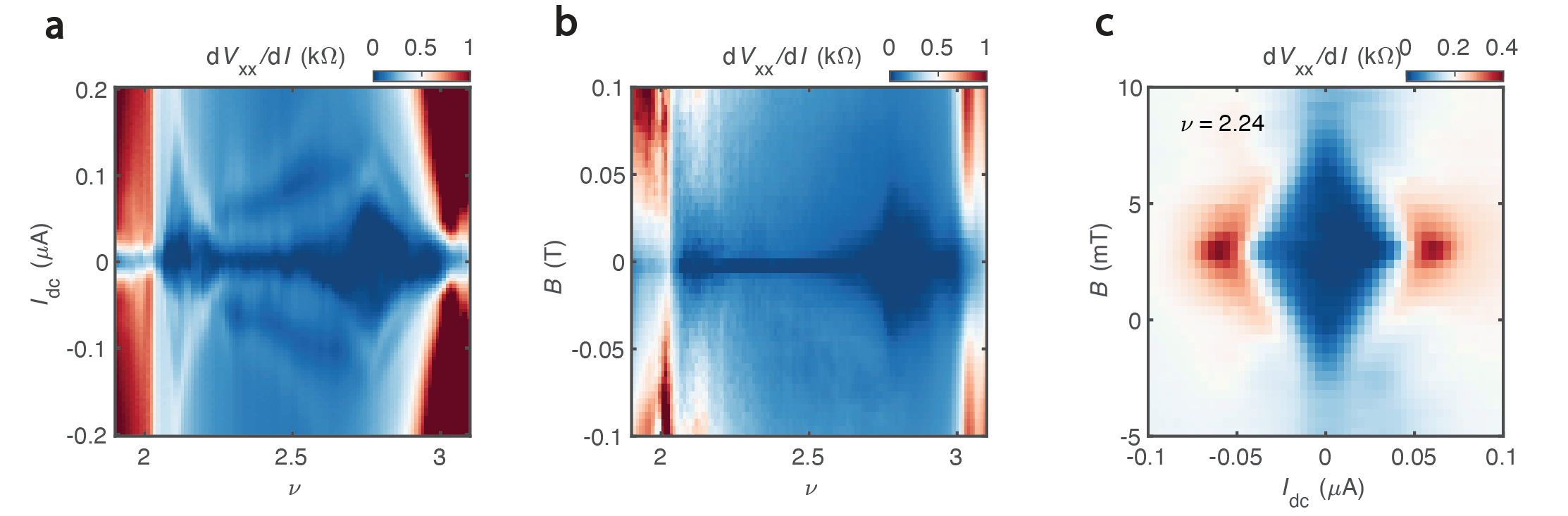}
  \caption{\textbf{Signature of double-dome superconductivity on the electron side.} \textbf{a}, $dV_{xx}/dI$ versus $\nu$ and $I_{dc}$ on the electron side. Critical current is in the range of $100$~nA, and superconductivity is suppressed near $\nu=2.4$. \textbf{b}, $dV_{xx}/dI$ as a function of $\nu$ and $B$. Superconductivity is weak, with a maximum critical magnetic field of around $50$~mT. A suppression of superconductivity is visible near $\nu=2.4$. \textbf{c}, $dV_{xx}/dI$ versus $I_{dc}$ and $B$ shows a clear `diamond' feature. The offset of the center of the magnetic field is due to the residual magnetic field in the magnet.}
  \label{sub8}
\end{figure*}

\begin{figure*}
\renewcommand{\thefigure}{S9}
  \centering
  \includegraphics[width= 0.684\textwidth]{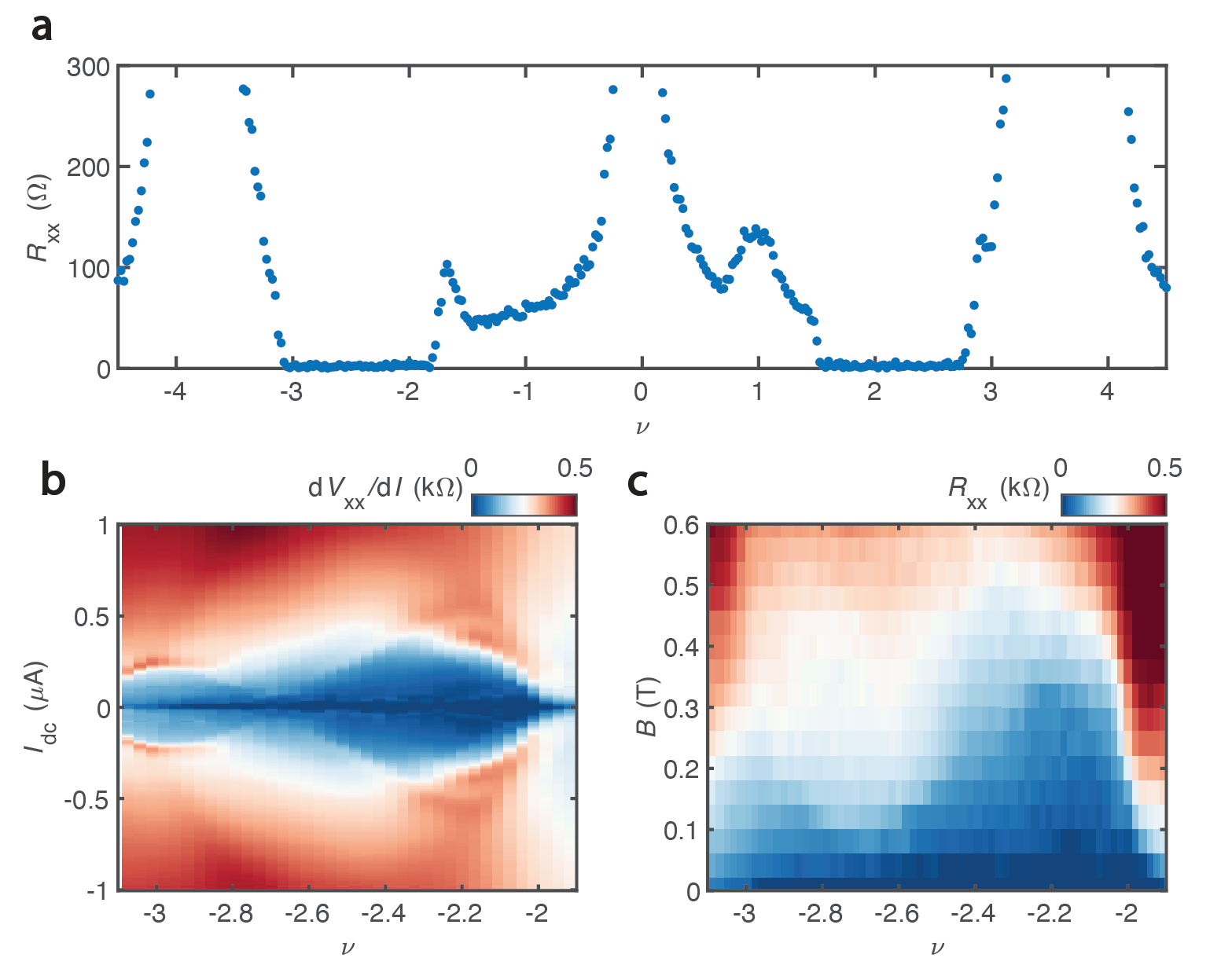}
  \caption{\textbf{Double-dome superconductivity in another TTG device with alternating twist angle $\theta=1.51^{\circ}$.} \textbf{a}, $R_{xx}$ as a function of bottom gate voltage $V_{bg}$ which is already converted into moiré filling $\nu$. Superconductivity is observed both on the hole side and the electron side. \textbf{b}, $dV_{xx}/dI$ versus $I_{dc}$ and $\nu$ on the hole side. Superconductivity is weakened around $\nu=-2.6$ and shows a double-dome behavior. \textbf{c}, Magnetic field dependence of $R_{xx}$ also shows double-dome superconductivity.}
  \label{sub9}
\end{figure*}

\end{document}